\documentclass[10pt,aps,prd,twocolumn,superscriptaddress,showpacs,nofootinbib,showkeys]{revtex4}
\pdfoutput=1

\usepackage{graphicx}
\usepackage[colorlinks=false,linktocpage=true]{hyperref}

\begin{document}
 
\title{Testing viscous and anisotropic hydrodynamics in an exactly solvable case}

\author{Wojciech Florkowski} 
\affiliation{Institute of Physics, Jan Kochanowski University, PL-25406~Kielce, Poland} 
\affiliation{The H. Niewodnicza\'nski Institute of Nuclear Physics, Polish Academy of Sciences, PL-31342 Krak\'ow, Poland}

\author{Radoslaw Ryblewski} 
\affiliation{The H. Niewodnicza\'nski Institute of Nuclear Physics, Polish Academy of Sciences, PL-31342 Krak\'ow, Poland} 

\author{Michael Strickland} 
\affiliation{Department of Physics, Kent State University, Kent, OH 44242 United States}
\affiliation{Frankfurt Institute for Advanced Studies\\
D-60438, Frankfurt am Main, Germany}

\date{\today}

\begin{abstract}
We exactly solve the one-dimensional boost-invariant Boltzmann equation in the relaxation time approximation for arbitrary shear viscosity. 
The results are compared with the predictions of viscous and anisotropic hydrodynamics. 
Studying different non-equilibrium cases and comparing the exact kinetic-theory results to the second-order viscous hydrodynamics results we find that recent formulations of second-order viscous hydrodynamics agree better with the exact solution than the standard Israel-Stewart approach. 
Additionally, we find that, given the appropriate connection between the kinetic and anisotropic hydrodynamics relaxation times, anisotropic hydrodynamics provides a very good approximation to the exact relaxation time approximation solution.
\end{abstract}



\maketitle 

\section{Introduction}
\label{sect:intro}

Since the seminal work of Israel and Stewart \cite{Israel:1976tn,Israel:1979wp} there has been a considerable amount of effort devoted to the development and application of relativistic viscous hydrodynamics  \cite{Muronga:2001zk,Muronga:2003ta,Baier:2006um,Romatschke:2007mq,Baier:2007ix,Dusling:2007gi,Luzum:2008cw,Song:2008hj,El:2009vj,Martinez:2009mf,PeraltaRamos:2010je,Denicol:2010tr,Denicol:2010xn,Schenke:2010rr,Schenke:2011tv,Bozek:2011wa,Niemi:2011ix,Niemi:2012ry,Bozek:2009dw,Bozek:2012qs,Denicol:2012cn,Jaiswal:2013npa}.  
Although the original developments of Israel and Stewart were primarily intended for application to astrophysical systems, recent developments have focused on the application of relativistic viscous hydrodynamics to the modeling of the hot and dense matter created in ultrarelativistic heavy ion collisions.  
These systems are special in the sense that the hot and dense matter created in this manner is subject to rapid longitudinal expansion along the beam line direction.  
As a result, the viscous corrections to the ideal energy-momentum tensor can become large (particularly the corrections to the longitudinal pressure).  

Because of these large corrections, many groups have been seeking methods to improve the description of the early-time dynamics of the hot and dense matter created in heavy ion collisions.  
These methods include, for example, complete second-order treatments \cite{Denicol:2012cn}, third-order treatments \cite{El:2009vj,Jaiswal:2013vta}, and a method dubbed anisotropic hydrodynamics which linearizes instead around a momentum-space anisotropic background \cite{Florkowski:2010cf,Martinez:2010sc,Ryblewski:2010bs,Martinez:2010sd,Ryblewski:2011aq,Martinez:2012tu,Ryblewski:2012rr,Ryblewski-0954-3899-40-9-093101,Florkowski:2012ax,Florkowski:2012as}.  
In order to assess the efficacy of these different approaches it would be nice to have an exactly solvable case with which to compare the various approximation schemes.

In this paper we present details concerning the development of ideas introduced in Ref.~\cite{Florkowski:2013lza} in which we presented a solution to the 0+1d Boltzmann equation in the relaxation time approximation (RTA). 
We  follow the method used in \cite{Baym:1984np,Baym:1985tna} (see also \cite{Heiselberg:1995sh,Wong:1996va}) to exactly solve the one-dimensional boost-invariant kinetic equation with the collision term treated in the relaxation-time approximation.  
We extend the exact solution to the case that the relaxation time, $\tau_{\rm eq}$, depends on proper time. 
Our results are then compared to second-order viscous hydrodynamics approximations \cite{Israel:1976tn,Israel:1979wp,Muronga:2001zk,Muronga:2003ta,Baier:2006um,Baier:2007ix,
Romatschke:2007mq,Dusling:2007gi,Luzum:2008cw,Song:2008hj,El:2009vj,PeraltaRamos:2010je,Denicol:2010tr,Denicol:2010xn,Schenke:2010rr,Schenke:2011tv,Bozek:2009dw,Bozek:2011wa,Niemi:2011ix,Niemi:2012ry,Bozek:2012qs,Denicol:2012cn,Jaiswal:2013npa} and the spheroidal anisotropic hydrodynamics approximation \cite{Florkowski:2010cf,Martinez:2010sc,Ryblewski:2010bs,Martinez:2010sd,Ryblewski:2011aq,Martinez:2012tu,Ryblewski:2012rr,Ryblewski-0954-3899-40-9-093101,Florkowski:2012ax,Florkowski:2012as}. 

The comparison of the exact kinetic theory result with the viscous and anisotropic hydrodynamics approximations allows us to characterize the effectiveness of various hydrodynamic approaches and to unambiguously establish the correct value of the shear viscosity coefficient in RTA. 
We find that the recent formulations of second-order viscous hydrodynamics \cite{Denicol:2012cn,Jaiswal:2013npa} better reproduce the results of the kinetic theory than the standard second-order (Israel-Stewart) approach. 
Additionally, we compare the predictions of the kinetic theory with the results of anisotropic hydrodynamics. 
We find very good agreement between the anisotropic hydrodynamics approximation and the exact solution, provided that the relaxation times used in the kinetic equation and anisotropic hydrodynamics are properly matched.

In Refs.~\cite{Martinez:2010sc,Martinez:2010sd,Martinez:2012tu} the equations of anisotropic hydrodynamics were derived from the kinetic theory with the collision term treated in RTA. 
This approach used the zeroth and the first moments of the kinetic equation. 
In addition, the distribution function was assumed to have the Romatschke-Strickland form (RSF) \cite{Romatschke:2003ms}. 
If the system is close to equilibrium, this formulation has a direct connection to the standard second-order viscous hydrodynamics --- the parameters of anisotropic hydrodynamics are connected with the kinetic coefficients. 
In this work, we reanalyze this connection and, compared to the original Martinez and Strickland paper \cite{Martinez:2010sc}, find a modified relation between the shear viscosity and the relaxation time which leads to a better agreement between the exact results of the exact kinetic theory solution and anisotropic hydrodynamics.

In addition, we establish the relationship between the relaxation time in the anisotropic hydrodynamics approximation and the exact case.  
By analyzing systems which are close to equilibrium, we analytically prove that there is a factor of two difference between the relaxation time in these two cases.
The need for a modification of the relations between the parameters of anisotropic hydrodynamics and the exact solution may be traced back to the use of RSF which restricts the distribution function to a spheroidal form. 
Once this factor of two is taken into account, we find that there is very good agreement between the anisotropic hydrodynamics approximation and the exact solution, with the agreement becoming better as the relaxation time decreases.  
Additionally, for far-from-equilibrium systems we demonstrate that the scale for the relaxation time is set by the transverse-momentum scale, $\Lambda$.

We  show that with the proper matching between the relaxation times used in the kinetic theory and anisotropic hydrodynamics approaches, one finds an excellent agreement between the two approaches.  
It is somewhat surprising that already at the leading order of the anisotropic hydrodynamics approximation, one obtains agreement with the exact results which is at the level achieved only in the second order of viscous hydrodynamics.  
This observation is demonstrated for a variety of different initial temperatures, initial momentum-space anisotropies, and values of the shear viscosity to entropy density ratio.

Our study is complementary to studies based on the AdS/CFT correspondence \cite{Mateos:2011ix,Mateos:2011tv,Chernicoff:2012iq,Chernicoff:2012gu,Gahramanov:2012wz,Heller:2011ju,Heller:2012je}. 
In both cases one checks how a system which is governed by specific non-equilibrium dynamics approaches the viscous hydrodynamic limit. 
Although the underlying model employed herein is rather simple due to the restriction to 0+1d dynamics and RTA, it allows one to easily study the system for different values of the shear viscosity to entropy ratio.  
Additionally, due to the simplicity of this toy-model the exact solutions can be obtained to arbitrary numerical accuracy allowing for precision tests.

The structure of the paper is as follows.  
In Sec.~\ref{sect:kineq} we introduce the kinetic equation to be solved and define a convenient set of variables which can be used in studies of the 0+1d Boltzmann equation.  
In Sec.~\ref{sect:enmomcon} we demonstrate how to compute moments of the Boltzmann equation using these variables.  
In Sec.~\ref{sect:sol} we write down an integral equation which exactly solves the 0+1d RTA kinetic equations.  
We then discuss how to numerically solve this equation and extract the components of the energy momentum tensor, the number density, and the entropy density. 
In Secs.~\ref{sect:viscous} and \ref{sect:aniso} we compare the exact solution to the kinetic equations to first and second order viscous hydrodynamics and anisotropic hydrodynamics.
In Sec.~\ref{sect:latetime} we derive the relationship between the shear viscosity and relaxation time by making use of asymptotic expansions of the dynamical equations.  
In Sec.~\ref{sect:conclusions} we conclude and give an outlook for the future.  
We collect some relations and properties of special functions which appear in App.~\ref{app:H}.
Finally, in App.~\ref{app:lambda1} we assess the effect of the inclusion of the full set of conformal terms in the Israel-Stewart evolution equations.

\section{Kinetic equation}
\label{sect:kineq}

\subsection{Relaxation-time approximation}
\label{sect:rta}

Our approach is based on the simple form of the kinetic equation 
\begin{equation}
 p^\mu \partial_\mu  f(x,p) =  C[ f(x,p)].
\label{kineq}
\end{equation}
Here $f(x,p)$ is the parton phase-space distribution function which depends on the parton space-time coordinates $x$ and momentum $p$, and $C$ is the collision term treated in RTA,
\begin{eqnarray}
C[f] = \frac{p \cdot u}{\tau_{\rm eq}} \left( f^{\rm eq}-f \right) .
\label{col-term}
\end{eqnarray}
The quantity $\tau_{\rm eq}$ is the relaxation time which can depend on proper time. 
The equilibrium distribution function $f^{\rm eq}$ may be taken to be a Bose-Einstein, Fermi-Dirac, or Boltzmann distribution which depends on $p \cdot u$ and the isotropic temperature $T$.  
In order to simplify some intermediate steps one can, without loss of generality, assume that $f$ is given by a Boltzmann distribution
\begin{eqnarray}
f^{\rm eq} = \frac{2}{(2\pi)^3} \exp\left(- \frac{p \cdot u}{T} \right).
\label{Boltzmann}
\end{eqnarray}
The factor of 2 above accounts for spin degeneracy.\footnote{Degeneracies such as color will be taken into account with an additional overall degeneracy factor $g_0$.}
The temperature $T$ will be obtained via the Landau matching condition which demands that the energy density calculated from the distribution function $f$ is equal to the energy density determined from the equilibrium distribution, $f^{\rm eq}$. 
The quantity $u^\mu$ represents the four-velocity of the local rest frame of the matter.  

We emphasize that, except for the entropy density, all results obtained are independent of the assumed form of the underlying distribution function up to trivial rescalings.  
For the entropy density, one need only change the relationship between the entropy density and the underlying distribution function.
We also note that, in the general case, the temperature $T$ above should be treated as an effective temperature related to the fourth root of the energy density and, therefore, it can be seen as an alternative measure of the energy density.
Only if the system is close to equilibrium does the definition of $T$ coincide with the standard concept of temperature.

The use of the simple form of the kinetic equation given in Eq.~(\ref{kineq}) is motivated by the fact that there are many results obtained within this approximation allowing us to make comparisons with other approaches. 
In particular, there exist several calculations of the kinetic coefficients in RTA, for example, see  \cite{Anderson1974466,Czyz:1986mr,Dyrek:1986vv,Cerc:2002,Romatschke:2011qp}.  
In addition, as we will demonstrate below, in this simple case it is possible to solve the kinetic equation exactly to arbitrary numerical precision using straightforward numerical algorithms.

In equilibrium, for massless particles obeying classical statistics one may use the following expressions for particle density, entropy density, energy density, and pressure
\begin{eqnarray}
n_{\rm eq} = \frac{2 g_0 T^3}{\pi^2}, \quad
{\cal S}_{\rm eq} = \frac{8 g_0 T^3}{\pi^2},
\nonumber \\
{\cal E}_{\rm eq} = \frac{6 g_0 T^4}{\pi^2}, \quad
{\cal P}_{\rm eq} = \frac{2 g_0 T^4}{\pi^2},
\label{eq-therm}
\end{eqnarray}
where $g_0$ is the degeneracy factor counting all internal degrees of freedom except for spin (the spin degeneracy equals 2).  In what follows we make use of the relation ${\cal E}_{\rm eq} = 3 {\cal P}_{\rm eq}$ when a specification of the equilibrium equation of state is required.

\subsection{Boost-invariant variables}
\label{sect:boostinvvar}

In the case of one-dimensional boost-invariant expansion, all scalar functions of time and space depend only on the proper time $\tau = \sqrt{t^2-z^2}$. 
In addition, the hydrodynamic flow $u^\mu$ has the following form \cite{Bjorken:1982qr}
\begin{eqnarray}
u^\mu = \left(\frac{t}{\tau},0,0,\frac{z}{\tau}\right) .
\label{U}
\end{eqnarray}
The phase-space distribution function $f(x,p)$ behaves like a scalar under Lorentz transformations. 
The requirement of boost invariance implies that in this case $f(x,p)$ may depend only on three variables: $\tau$, $w$ and $\vec{p}_T$ \cite{Bialas:1984wv,Bialas:1987en}. 
The boost-invariant variable $w$ is defined by 
\begin{equation}
w =  tp_L - z E \, .
\label{w}
\end{equation}
With the help of $w$ and $\vec{p}_T$ we define 
\begin{equation}
v(\tau,w,p_T) = Et-p_L z = 
\sqrt{w^2+\left( m^2+\vec{p}_T^{\,\,2}\right) \tau^2} \, .  
\label{v}
\end{equation}
From (\ref{w}) and (\ref{v}) one can easily find the energy and the longitudinal momentum of a particle 
\begin{equation}
E= p^0 = \frac{vt+wz}{\tau^2} \, ,\quad p_L=\frac{wt+vz}{\tau^2} \, .  
\label{p0p3}
\end{equation}
The momentum integration measure in phase-space is 
\begin{equation}
dP = 2 \, d^4p \, \delta \left( p^2-m^2\right) \theta (p^0)
=\frac{dp_L}{p^0}d^2p_T =\frac{dw}vd^2p_T \, .  
\label{dP}
\end{equation}
In the following we shall consider massless partons and set masses equal to zero, $m=0$.

\subsection{Boost-invariant form of the kinetic equation}
\label{sect:binvkineq}

Using the boost-invariant variables introduced in the previous Section one finds 
\begin{eqnarray}
p^\mu \partial_\mu f = 
\frac{v}{\tau} \frac{\partial f}{\partial \tau}, \quad 
p \cdot u = \frac{v}{\tau} \, .
\label{binvterms}
\end{eqnarray}
Using Eq.~(\ref{binvterms}) in Eq.~(\ref{kineq}) and simplifying, one finds 
\begin{eqnarray}
\frac{\partial f}{\partial \tau}  &=& 
\frac{f^{\rm eq}-f}{\tau_{\rm eq}} \, ,
\end{eqnarray} 
where the equilibrium distribution function may be written as
\begin{eqnarray}
f^{\rm eq}(\tau,w,p_T) =
\frac{2}{(2\pi)^3} \exp\left[
- \frac{\sqrt{w^2+p_T^2 \tau^2}}{T(\tau) \tau}  \right].
\end{eqnarray}
In the following we assume that 
$f(\tau,w,\vec{p}_T)$ 
is an even function of $w$ and depends only on the magnitude of the transverse momentum $\vec{p}_T$,
\begin{eqnarray}
f(\tau,w,p_T) = f(\tau,-w,p_T) \, .
\label{symofg}
\end{eqnarray}

\section{Moments of the kinetic equation}
\label{sect:enmomcon}

In this Section we detail how to calculate the moments of the kinetic equation using the coordinates introduced in the previous Section.  
In addition, we discuss the application of dynamical Landau matching which results from the requirement of energy conservation.

\subsection{Zeroth moment -- parton number current}
\label{sect:partnumber}

The zeroth moment of the kinetic equation (\ref{kineq}) leads to the equation
\begin{eqnarray}
\frac{dn}{d\tau} + \frac{n}{\tau} = 
\frac{n^{\rm eq}-n}{\tau_{\rm eq}} \, ,
\label{dndt}
\end{eqnarray}
where the parton density (measured in the local rest frame) equals
\begin{eqnarray}
n(\tau) &=&  g_0 \int dP\, p \cdot u f(\tau,w,p_T) \, , \nonumber \\
&=& \frac{g_0}{\tau} \int dP\, v\, f(\tau,w,p_T) \, .
\end{eqnarray}
We note that the parton number is not conserved in RTA. 
This is in agreement with the expectations that partons (gluons) are produced at the early stages of the collisions.   
In the approaches where the parton density is proportional to the entropy density, the right-hand side of (\ref{dndt}) is proportional to the entropy source term.
Note, however, that it is possible to enforce baryon number conservation in the quark sector in
the anisotropic hydrodynamics framework \cite{Florkowski:2012as}.\footnote{To conserve parton number one may introduce an effective chemical potential in an analogous way as one introduces the effective temperature.}
Herein we will ignore the distinction between quarks and gluons and simply treat the system as partons with the same relaxation time and bulk properties.

\subsection{First moment -- energy-momentum tensor}
\label{sect:enmomten}

The first moment of the left-hand side of Eq.~(\ref{kineq}) defines the divergence of the energy-momentum tensor
\begin{equation}
T^{\mu\nu}(\tau) = g_0 \int dP \, p^\mu p^\nu f(\tau,w,p_T) \, . \label{Tmunu1}
\end{equation}
Using the symmetry properties (\ref{symofg}) we may rewrite (\ref{Tmunu1}) in the form 
\cite{Florkowski:2010cf,Martinez:2012tu}
\begin{equation}
T^{\mu\nu} = ({\cal E} + {\cal P}_T) u^\mu u^\nu - {\cal P}_T g^{\mu\nu} + ({\cal P}_L-{\cal P}_T) z^\mu z^\nu \, , 
\label{Tmunu2}
\end{equation}
where
\begin{eqnarray}
{\cal E}(\tau) &=& \frac{g_0}{\tau^2}\,
\int dP \, v^2\,  f(\tau,w,p_T) \, , \nonumber \\
{\cal P}_L(\tau) &=& \frac{g_0}{\tau^2}\,
\int dP \, w^2\,  f(\tau,w,p_T) \, , \nonumber \\
{\cal P}_T(\tau) &=& \frac{g_0}{2}\,
\int dP \, p_T^2\, f(\tau,w,p_T) \, ,
\label{epsandpres}
\end{eqnarray}
and
\begin{eqnarray}
z^\mu = \left(\frac{z}{\tau},0,0,\frac{t}{\tau}\right) \, ,
\label{V}
\end{eqnarray}
is a four-vector orthogonal to $u^\mu$ which, in the local rest frame, is the $z$-direction of the coordinate system.

The energy-momentum conservation law for the system of partons has the form
\begin{equation}
\partial _\mu T^{\mu \nu }(x)=0 \,.
\label{enmomcon1}
\end{equation}
For one dimensional systems, the four equations implied by Eq.~(\ref{enmomcon1}) are reduced to the single equation
\begin{eqnarray}
\frac{d{\cal E}}{d\tau}=
- \frac{{\cal E}+{\cal P}_L}{\tau} \, .
\label{enmomcon2}
\end{eqnarray}

We note that the structure of the energy-momentum tensor (\ref{Tmunu2}) with (\ref{epsandpres}) and (\ref{V}) is typical for an anisotropic system. 
In the framework of anisotropic hydrodynamics one solves Eqs.~(\ref{dndt}) and (\ref{enmomcon2}) with the assumption that all bulk properties (such as $n$, ${\cal E}$, ${\cal P}_L$, and ${\cal P}_T$) can be expressed in terms of two independent variables. 
These variables can be chosen, for example, to be the longitudinal and transverse pressures, the entropy density ${\cal S}$ and the anisotropy parameter $x$ which is related to the momentum-space ellipticity of the  distribution function \cite{Florkowski:2010cf}, or the transverse momentum scale $\Lambda$ and the anisotropy parameter $\xi = x-1$  \cite{Martinez:2010sc}. 
Within kinetic theory, Eqs.~(\ref{dndt}) and (\ref{enmomcon2}) are automatically fulfilled if the distribution function satisfies the kinetic equation (\ref{kineq}).

\subsection{Landau matching}
\label{sect:LM}

Equation (\ref{enmomcon1}) is satisfied at any proper time if the energy densities calculated with the distribution functions $f$ and $f^{\rm eq}$  are equal, namely
\begin{eqnarray}
{\cal E}(\tau) &=& \frac{g_0}{\tau^2}\,
\int dP \, v^2\,  f(\tau,w,p_T) \, ,
\nonumber \\ 
&=& \frac{g_0}{\tau^2}\,
\int dP \, v^2\,  f^{\rm eq}(\tau,w,p_T) \, ,
\nonumber \\
&=& 
 \frac{6 g_0 T^4(\tau)}{\pi^2} \, . \label{LM1}
\end{eqnarray}
The last line above defines the {\it effective temperature} $T(\tau)$ that should be used in the background distribution function.

\section{Solutions of the kinetic equation}
\label{sect:sol}

In this Section we introduce the general structure of solutions of the kinetic equation (\ref{kineq}) and present numerical solutions for different initial conditions. 
The latter are characterized by the initial momentum anisotropy $x_0=1+\xi_0$, the initial effective temperature $T_0$, and the initial proper time $\tau_0$. 
The time dependence of the physical quantities such as energy density or the two pressures depends on the specific form of the relaxation time. 
The results presented in this Section will be used to make comparisons with viscous and anisotropic hydrodynamics in the next Sections.

\subsection{Formal structure of solutions and damping function}
\label{sect:formsol}

The formal solution of the kinetic equation (\ref{kineq}) has the form
\begin{eqnarray}
f(\tau,w,p_T) &=& D(\tau,\tau_0) f_0(w,p_T)  \label{solG} \\
&+&  \int_{\tau_0}^\tau \frac{d\tau^\prime}{\tau_{\rm eq}(\tau^\prime)} \, D(\tau,\tau^\prime) \, 
f^{\rm eq}(\tau^\prime,w,p_T) \, ,  \nonumber
\end{eqnarray}
where we have introduced the damping function
\begin{eqnarray}
D(\tau_2,\tau_1) = \exp\left[-\int\limits_{\tau_1}^{\tau_2}
\frac{d\tau^{\prime\prime}}{\tau_{\rm eq}(\tau^{\prime\prime})} \right] .
\end{eqnarray}
For \mbox{$\tau=\tau_0$} the distribution function $f$ is reduced to the initial distribution function, $f_0$. 

The damping function $D(\tau_2,\tau_1)$ has the following properties: $D(\tau,\tau) = 1$, $ D(\tau_3,\tau_2) D(\tau_2,\tau_1) = D(\tau_3,\tau_1)$, and
\begin{eqnarray}
\frac{\partial D(\tau_2,\tau_1)}{\partial \tau_2}
= -\frac{D(\tau_2,\tau_1)}{\tau_{\rm eq}(\tau_2)} \, .
\label{Dprop}
\end{eqnarray}
The equilibration time in our approach may be an arbitrary function of the proper time, $\tau_{\rm eq}=\tau_{\rm eq}(\tau)$. 
For the exact solution we use the relation
\begin{eqnarray}
\tau_{\rm eq}(\tau) = \frac{5 {\bar \eta}}{T(\tau)},
\label{taueq}
\end{eqnarray}
where ${\bar \eta} \equiv \eta/{\cal S}$ is the ratio of the shear viscosity to entropy density.
We will assume that $\bar\eta$ is time-independent in all results that follow.  
We return to the discussion of the relationship between $\bar\eta$ and $\tau_{\rm eq}$ in Secs.~\ref{sect:first} and \ref{subsec:vhydroasymp}.

In the numerical calculations we use the values
\begin{eqnarray}
{\bar \eta} \in \left\{ \frac{1}{4\pi} \, , \, \frac{3}{4\pi} \, ,\, \frac{10}{4\pi} \right\} \, .
\label{etabars}
\end{eqnarray}
The first two values on the right hand side of (\ref{etabars}) cover the viscosity range extracted to date from the hydrodynamic analyses of relativistic heavy-ion collisions studied at RHIC and the LHC.  
The last value is on the order expected by leading log perturbative results extrapolated to RHIC and LHC energies.

Applying the Landau matching condition (\ref{LM1}) directly to the formal solution (\ref{solG}) one finds
\begin{eqnarray}
T^4(\tau) &=& D(\tau,\tau_0) \frac{\pi^2 {\cal E}^0(\tau)}{6g_0}  \label{solT} \\
&& + \int_{\tau_0}^\tau \frac{d\tau^\prime}{2 \tau_{\rm eq}(\tau^\prime)} \, D(\tau,\tau^\prime) \, 
T^4(\tau^\prime) {\cal H}\left(\frac{\tau^\prime}{\tau}\right).  \nonumber
\end{eqnarray}
Here ${\cal E}^0(\tau)$ denotes the weighted integral over the initial distribution function $f_0$,
\begin{eqnarray}
{\cal E}^0(\tau) &=& \frac{g_0}{\tau^2}\,
\int dP \, v^2\,  f_0(w,p_T). \label{eps0} 
\end{eqnarray}
We stress that the time dependence of ${\cal E}^0(\tau)$ is induced not only by the term $1/\tau^2$ but by the time dependence of $v$ as well. The initial energy density is given by
\begin{eqnarray}
{\cal E}_0 = {\cal E}^0(\tau_0)
= \frac{6 g_0 T_0^4}{\pi^2}.
\end{eqnarray}
The function ${\cal H}$ appearing in (\ref{solT}) may be expressed in terms of the function ${\cal R}$ defined in Refs.~\cite{Martinez:2010sc,Martinez:2010sd,Martinez:2012tu}, namely
\begin{eqnarray}
{\cal H}\left( y \right) &=& 2 \, 
{\cal R}\left(\frac{1}{y^2}-1\right),
\end{eqnarray}
where ${{\cal R}(z) = \frac{1}{2} \big[ (1+z)^{-1} + \arctan\!\big(\sqrt{z}\big)/\sqrt{z} \big]}$.  We give more details concerning the ${\cal H}$ and ${\cal R}$ functions in App.~\ref{app:H}. 

\subsection{Initial distributions}
\label{sect:initdistr}

\subsubsection{Romatschke-Strickland form}
\label{sect:RSform}

As our first option for the initial conditions we consider the Romatschke-Strickland form \cite{Romatschke:2003ms} with a Boltzmann distribution as the underlying isotropic distribution 
\begin{eqnarray}
f_0(w,p_T) &=& \frac{2}{(2\pi)^3}
\exp\left[
-\frac{\sqrt{(p\cdot u)^2 + \xi_0 (p\cdot z)^2}}{\Lambda_0} \, \right] \nonumber \\
&=& \frac{1}{4\pi^3}
\exp\left[
-\frac{\sqrt{(1+\xi_0) w^2 + p_T^2 \tau_0^2}}{\Lambda_0 \tau_0}\, \right].
\nonumber \\
\label{RS}
\end{eqnarray}
This reduces to an isotropic Boltzmann distribution if the anisotropy parameter $\xi_0=\xi(\tau_0)$ vanishes. In this case, the transverse momentum scale $\Lambda_0$ is equal to the system's initial temperature $T_0$. By direct calculation one obtains
\begin{eqnarray}
{\cal E}^0(\tau) &=& 
\frac{6 g_0 T_0^4}{\pi^2} \, 
\frac{{\cal H}
\left( \frac{\tau_0 }{\tau}\, x_0^{-1/2} \right)}
{{\cal H}\left(x_0^{-1/2}\right)} \, ,
\label{eps0tauRS}
\end{eqnarray}
where 
\begin{eqnarray}
x(\tau) = 1+\xi(\tau) \, ,
\label{iks}
\end{eqnarray}
and $x_0=x(\tau_0)$ denotes the initial value of the anisotropy parameter $x$.

\subsubsection{Gaussian distributions}
\label{sect:Gaussform}

As another option for the initial distribution function we consider an anisotropic Gaussian distribution of the form
\begin{eqnarray}
f_0(w,p_T) = A \exp\left[
-\frac{w^2}{C^2 \tau_0^2}-B^2 p_T^2 
\right],
\label{Gauss}
\end{eqnarray}
where the parameters $C$ and $B$ determine the width(s) of the distribution in momentum space and $A$ is an overall normalization. 
In this case the integral over the initial distribution function gives
\begin{eqnarray}
{\cal E}^0(\tau) &=& 
\frac{6 g_0 T_0^4}{\pi^2} \, 
\frac{{\cal H}
\left( \frac{ \tau_0 }{\tau} C B \right)}
{{\cal H}\left(C B\right)} \, .
\label{eps0tauG}
\end{eqnarray}
By comparing Eqs.~(\ref{eps0tauRS}) and (\ref{eps0tauG}) we see that the Romatschke-Strickland and Gaussian initial conditions lead to the same dynamic evolution equation for the effective temperature $T(\tau)$ via Eq.~(\ref{solT}) if one takes
\begin{eqnarray}
C B = x_0^{-1/2} = (1+\xi_0)^{-1/2}.
\end{eqnarray}
Consequently, in this work we will use the Romatschke-Strickland form from this point forward with the understanding that the evolution of the effective temperature is the same assuming the initial widths are chosen as described above.  
As a result, we can solve the following dynamical equation for the effective temperature 
\begin{eqnarray}
T^4(\tau) &=& D(\tau,\tau_0)T_0^4 \, 
\frac{{\cal H}
\left( \frac{\tau_0 }{\tau}\, x_0^{-1/2} \right)}
{{\cal H}\left(x_0^{-1/2}\right)} \label{solT1} \\
&&  + \int_{\tau_0}^\tau \frac{d\tau^\prime}{2 \tau_{\rm eq}(\tau^\prime)} \, D(\tau,\tau^\prime) \, 
T^4(\tau^\prime) {\cal H}\left(\frac{\tau^\prime}{\tau}\right).  \nonumber
\end{eqnarray}

We note that in the limit $C \to 0$ (at fixed $B$) or $\xi_0,x_0 \to \infty$ the initial distribution is very narrow in $w$, and the initial longitudinal pressure of the system vanishes, ${\cal P}_L(\tau_0) \to 0$. 
Such configurations naturally emerge in models of the very early stages of heavy ion collisions, for example, in the color glass condensate theory. 
The situation where the transverse pressure is larger than the longitudinal pressure corresponds to an ``oblate'' momentum-space distribution.

\begin{figure}[t]
\begin{center}
\includegraphics[angle=0,width=1.1\textwidth]{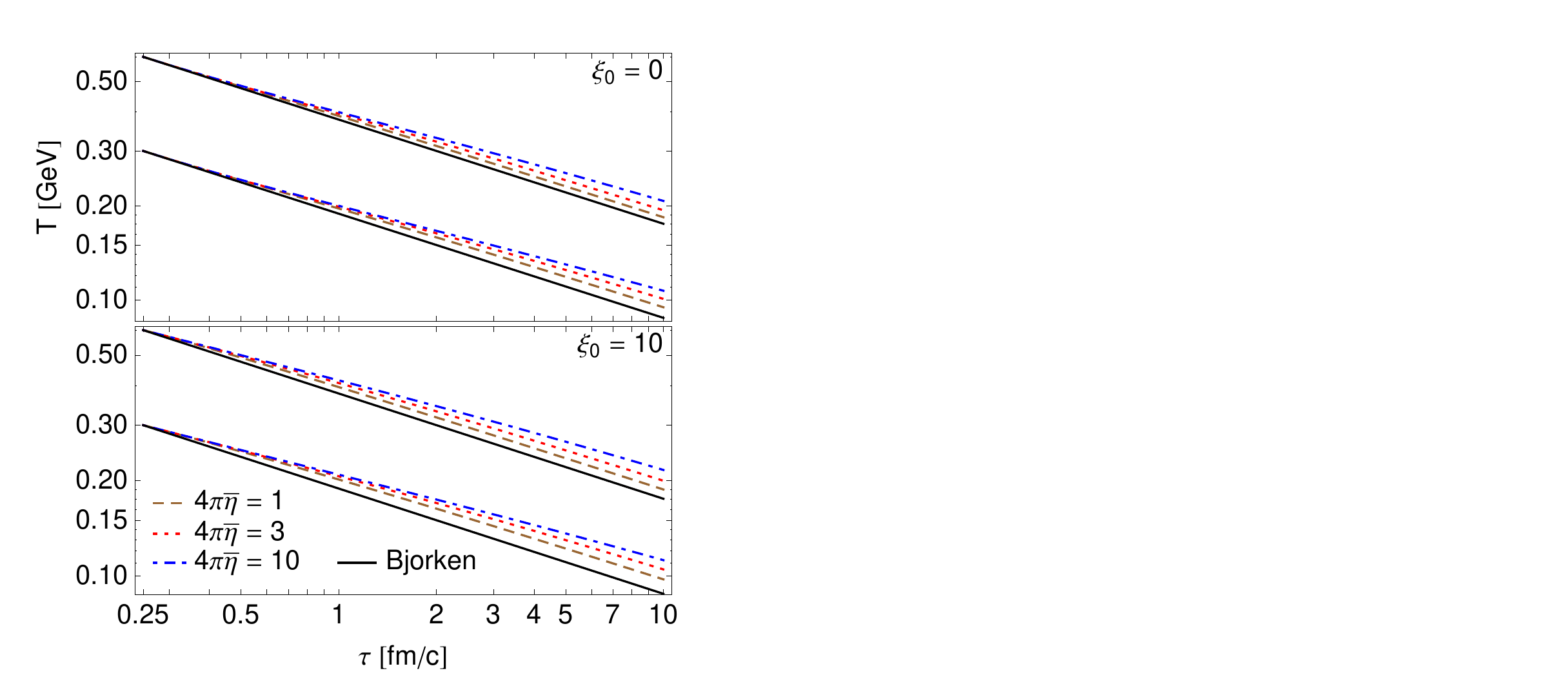}\end{center}
\caption{(Color online) Time dependence of the effective temperature $T(\tau)$ for two different values of the initial anisotropy: $\xi_0 = 0$ (upper panel) and $\xi_0=10$ (lower panel), and for two values of the initial temperature: \mbox{$T_0$ = 300 MeV} and \mbox{$T_0$ = 600 MeV}. The initial time $\tau_0$ = 0.25 fm/c. The dashed, dotted, and dashed-dotted lines correspond to different values of viscosity; $4 \pi {\bar \eta} =1, 3$ and 10, respectively. The solid black lines show the ideal Bjorken results corresponding to the limit $\tau_{\rm eq} \to 0$ and assuming that the initial distribution is an equilibrium distribution.}
\label{fig:T}
\end{figure}

\begin{figure}[t]
\begin{center}
\includegraphics[angle=0,width=1.1\textwidth]{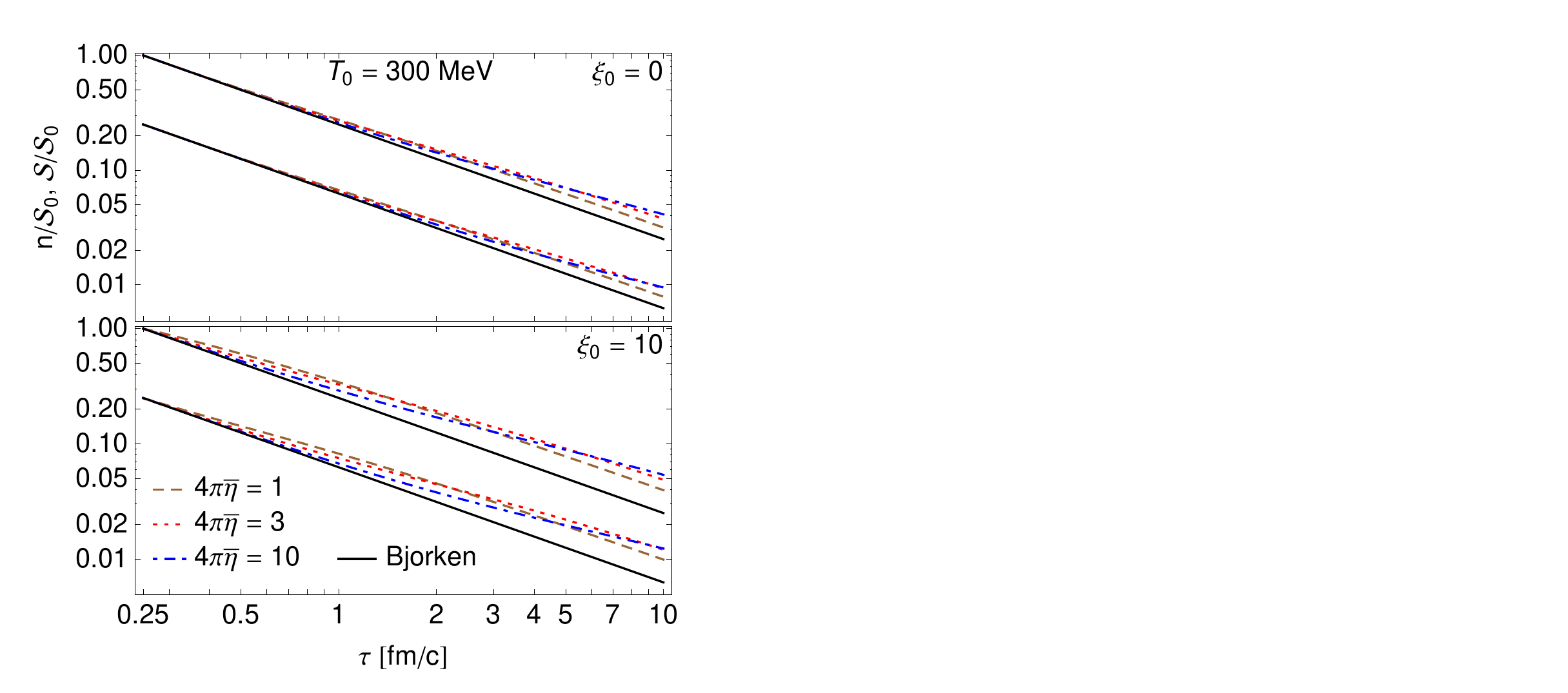}
\end{center}
\caption{(Color online) Time dependence of the parton number (lower lines) and entropy (upper lines) densities rescaled by the initial entropy density ${\cal S}_0$ for different values of viscosity. The solid black line shows the ideal Bjorken result where $n=n_0 \tau_0/\tau$ and ${\cal S}={\cal S}_0 \tau_0/\tau = 4 n_0 \tau_0/\tau$ . The initial temperature $T_0 =$ 300 MeV.}
\label{fig:ns_300}
\end{figure}

\begin{figure}[t]
\begin{center}
\includegraphics[angle=0,width=1.1\textwidth]{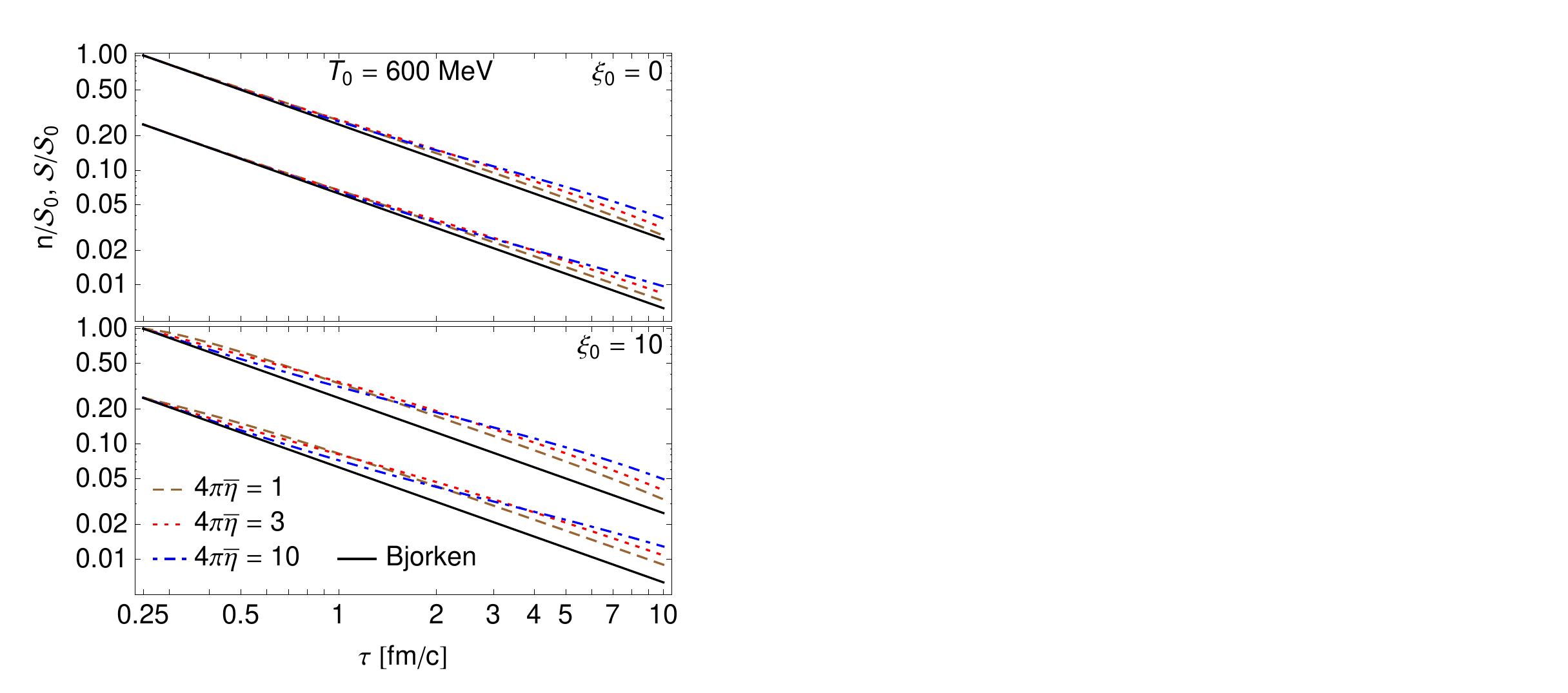}
\end{center}
\caption{(Color online) Same as Fig.~\ref{fig:ns_300} but for the initial effective temperature $T_0 =$ 600 MeV.}
\label{fig:ns_600}
\end{figure}

\subsection{Numerical method}
\label{sect:num}

Equation (\ref{solT1}) can be solved by the iterative method. 
We first use a trial function $T_a(\tau)$ and substitute it into the right-hand-side of Eq.~(\ref{solT1}). 
In this way the left-hand-side of (\ref{solT1}) defines the new temperature profile $T_b(\tau)$ which, in the next iteration, we treat as $T_a(\tau)$ and substitute into the right-hand-side of (\ref{solT1}). 
Repeating this procedure many times, we find a stable temperature profile which is invariant under further iterations. 
This method has been successfully used earlier, for example, in \cite{Banerjee:1989by}.

In Fig.~\ref{fig:T} we show the time dependence of the effective temperature $T(\tau)$ obtained from Eq.~(\ref{solT1}) for two different values of the initial anisotropy: $\xi_0 = 0$ (upper panel) and $\xi_0=10$ (lower panel), and for two values of the initial effective temperature: \mbox{$T_0$ = 300 MeV} and \mbox{$T_0$ = 600 MeV}. 
The initial time $\tau_0$ = 0.25 fm/c. 
The dashed, dotted, and dashed-dotted lines correspond to different values of viscosity; $4 \pi {\bar \eta} =1, 3$, and 10, respectively. 

The solid black lines in Fig.~\ref{fig:T} show the ideal Bjorken results corresponding to the limit $\tau_{\rm eq} \to 0$ and assuming that the initial distribution is an equilibrium distribution. 
With increasing viscosity we observe larger deviations from the ideal Bjorken solution.  This can be easily understood with the help of Eq.~(\ref{enmomcon2}) since larger values of viscosity imply smaller longitudinal pressure and hence a smaller decrease of the energy density with time.

\subsection{Parton and entropy densities}
\label{sect:observables1}

Once the effective temperature $T(\tau)$ is obtained, one may use it to find other bulk properties by performing the appropriate phase-space integrals. In particular, the parton density can be obtained via
\begin{eqnarray}
n(\tau) &=& \frac{2 g_0}{\pi^2} \left[ D(\tau,\tau_0) \, \Lambda_0^3 \, x_0^{-1/2} \frac{\tau_0}{\tau}  \right.
\nonumber \label{soln}\\
&& \left. + \int_{\tau_0}^\tau \frac{d\tau^\prime}{\tau_{\rm eq}(\tau^\prime)} \, D(\tau,\tau^\prime) \, 
T^3(\tau^\prime) \frac{\tau^\prime}{\tau} \right].
\end{eqnarray}
Using the Landau matching condition (\ref{LM1}) at $\tau = \tau_{0}$ one finds 
\begin{equation}
T_0^4 = \frac{1}{2}\,\Lambda_0^4 \,
{\cal H}\left(x_0^{-1/2}\right),
\label{LM0}
\end{equation}
which relates the initial values of $\Lambda_0$, $T_0$, and \mbox{$x_0 = 1+\xi_0$}. 

Assuming classical statistics, the entropy density can be calculated from the Boltzmann formula
\begin{eqnarray}
{\cal S}(\tau) &=& - g_0 \int dP \, p\cdot u\, f(\tau,w,p_T) 
\nonumber \label{sols}\\
&& \hspace{1cm} \times \left[\ln\left( 4 \pi^3 \, f(\tau,w,p_T)\right) - 1\right].
\label{sigmaoftau}
\end{eqnarray}
Here, the distribution function $f(\tau,w,p_T)$ is obtained from Eq.~(\ref{solG}). 
In equilibrium ${\cal S} = 4 n$ and the equilibrium pressure ${\cal P}_{\rm eq}$ is connected with the parton density by the well known relation ${\cal P}_{\rm eq} = n_{\rm eq} T$.

The time dependence of the parton and entropy densities extracted from the exact solution of Eq.~(\ref{kineq}) scaled by the initial entropy density ${\cal S}_0$ is shown in Figs.~\ref{fig:ns_300} and \ref{fig:ns_600} for $T_0 =$ 300 MeV and $T_0 =$ 600 MeV, respectively. 
In this case one observes  an interesting behavior: the finally produced entropy is larger in the cases with larger viscosity but this is only because entropy is produced in longer time intervals in such cases --- the initially produced entropy is larger when the viscosity is smaller. 
This non-monotonic behavior is different from that observed in the case of effective temperature (energy density) shown in Fig.~\ref{fig:T}.

\subsection{Longitudinal and transverse pressures}
\label{sect:observables2}

In a similar manner, one can obtain the longitudinal pressure
\begin{eqnarray}
{\cal P}_L(\tau) &=& \frac{6 g_0}{\pi^2} \left[ D(\tau,\tau_0) \,T_0^4  \, 
\frac{{\cal H}_L
\left( \frac{\tau_0 }{\tau}\, x_0^{-1/2} \right)}
{{\cal H}\left(x_0^{-1/2}\right)} \right.
\label{PLform} \\
&& \left. + \int_{\tau_0}^\tau \frac{d\tau^\prime}{2 \tau_{\rm eq}(\tau^\prime)} \, D(\tau,\tau^\prime) \, 
T^4(\tau^\prime) {\cal H}_L\left(\frac{\tau^\prime}{\tau}\right) \right], \nonumber
\end{eqnarray}
where ${\cal H}_L$ is defined by the expression\footnote{For more information about the functions ${\cal H}_T$ and ${\cal H}_L$ see Appendix \ref{app:H}.}
\begin{eqnarray}
{\cal H}_L(y) = y^2 \frac{d}{dy}
\left(\frac{{\cal H}(y)}{y} \right).
\end{eqnarray}
Replacing the function ${\cal H}_L$ by ${\cal H}_T$ where
\begin{eqnarray}
{\cal H}_T(y) = {\cal H}(y) - {\cal H}_L(y) \, ,
\end{eqnarray}
and dividing by two, one obtains the transverse pressure
\begin{eqnarray}
{\cal P}_T(\tau) &=& \frac{3 g_0}{\pi^2} \left[ D(\tau,\tau_0) \,T_0^4  \, 
\frac{{\cal H}_T
\left( \frac{\tau_0 }{\tau}\, x_0^{-1/2} \right)}
{{\cal H}\left(x_0^{-1/2}\right)} \right.
\label{PTform} \\
&& \left. + \int_{\tau_0}^\tau \frac{d\tau^\prime}{2 \tau_{\rm eq}(\tau^\prime)} \, D(\tau,\tau^\prime) \, 
T^4(\tau^\prime) {\cal H}_T\left(\frac{\tau^\prime}{\tau}\right) \right]. \nonumber
\end{eqnarray}
Equations~(\ref{PLform}) and (\ref{PTform}) will be used below when comparing the exact results obtained from the kinetic equation with the results obtained from the second-order viscous and anisotropic hydrodynamic approximations.

\begin{figure}[t]
\begin{center}
\includegraphics[angle=0,width=1\textwidth]{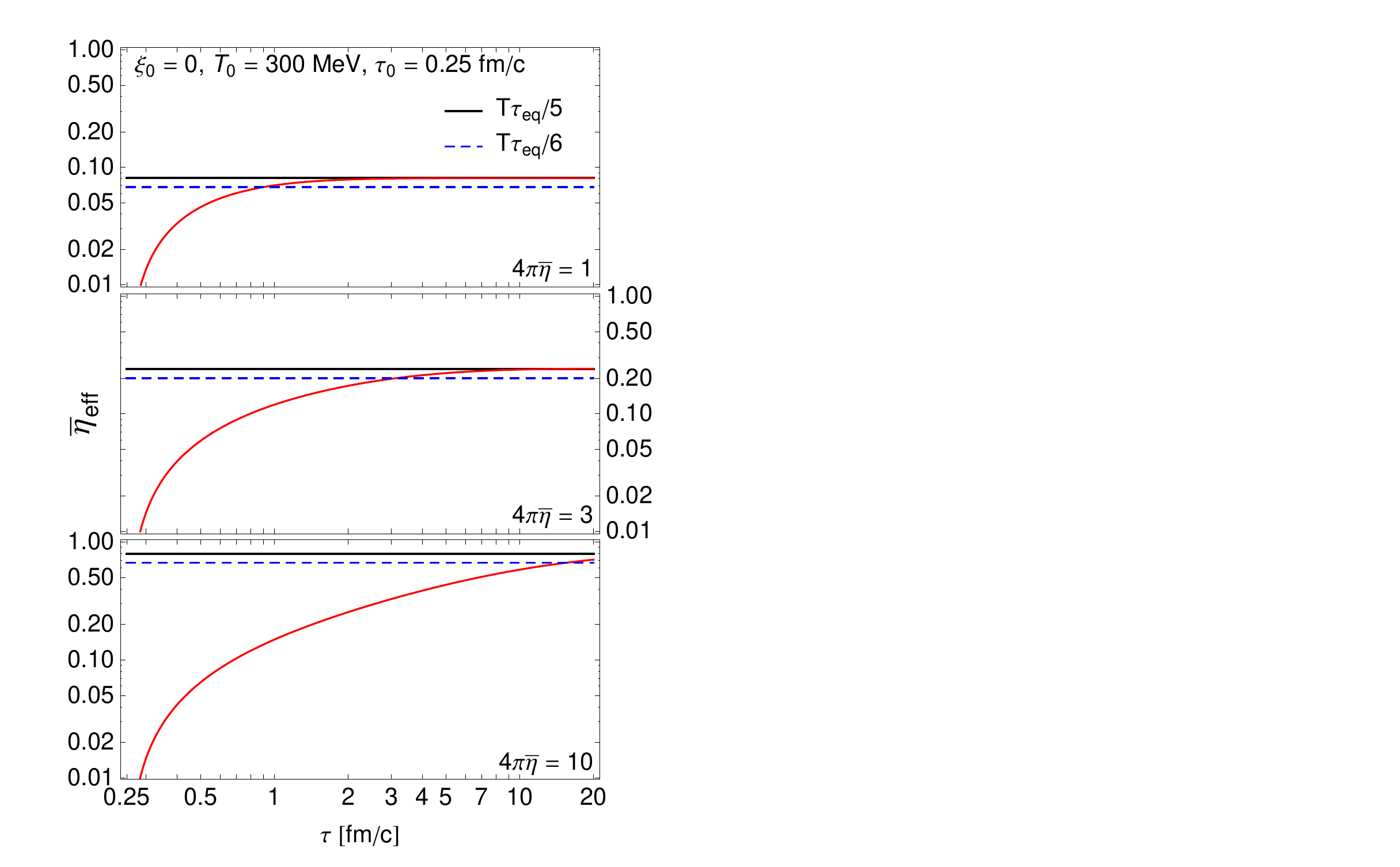}
\end{center}
\caption{(Color online) Time dependence of the effective shear viscosity to entropy density ratio ${\bar \eta}_{\rm eff}(\tau)$ obtained from Eq.~(\ref{eta-fo}) with $T(\tau)$ calculated from the kinetic equation (red lines). For comparison we show the two results for $\bar \eta$ which are discussed in the literature: ${\bar \eta} = T \tau_{\rm eq}/5$ (solid black lines) and ${\bar \eta} = T \tau_{\rm eq}/6$ (dashed blue lines) \cite{Cerc:2002}. One observes that for sufficiently large times our system is described by first-order hydrodynamics with
${\bar \eta} = T \tau_{\rm eq}/5$. The initial temperature is $T_0$ = 300 MeV.
}
\label{fig:etabar_300}
\end{figure}

\begin{figure}[t]
\begin{center}
\includegraphics[angle=0,width=1\textwidth]{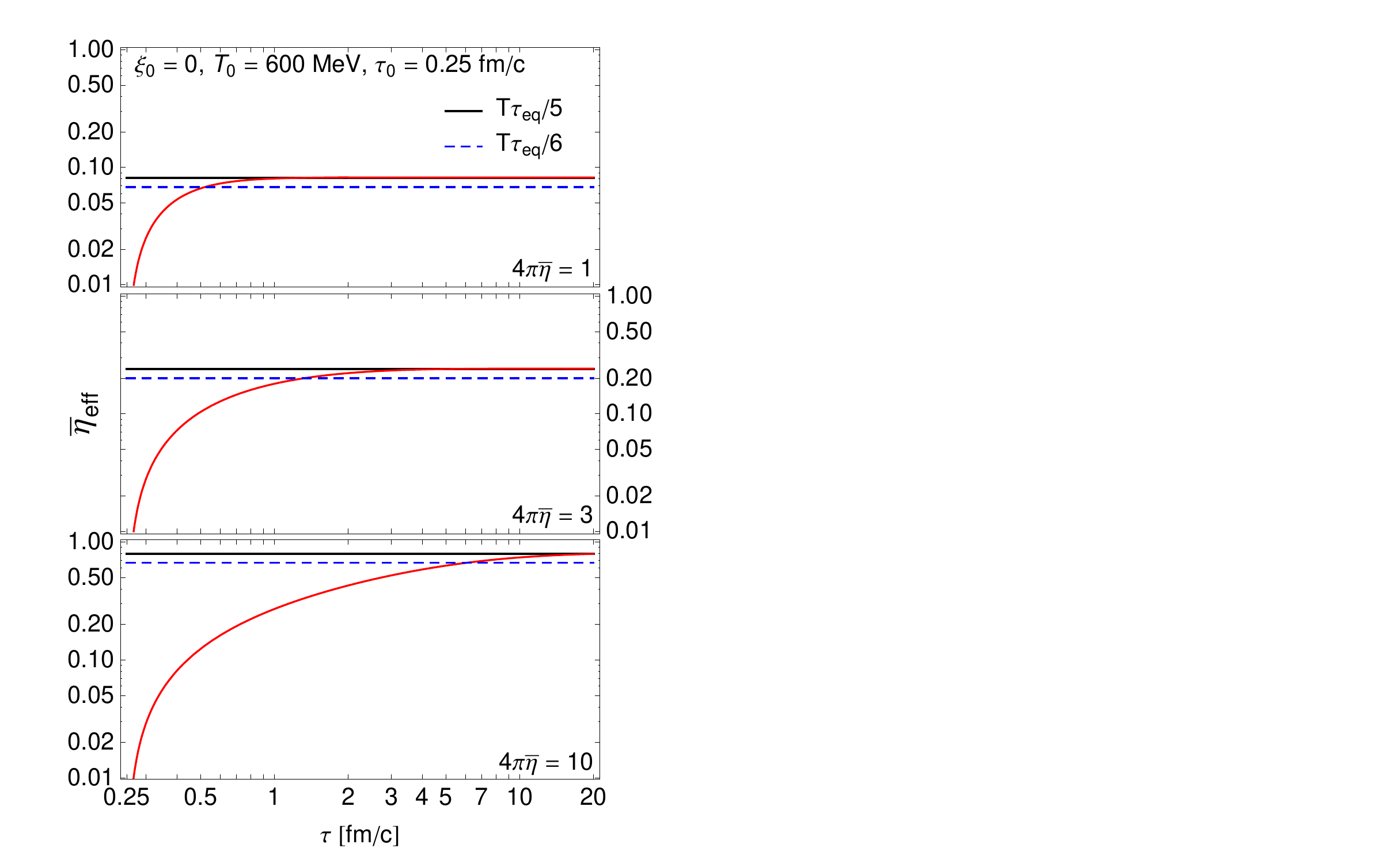}
\end{center}
\caption{(Color online) Same as Fig.~\ref{fig:etabar_300} but for $T_0$ = 600 MeV.
}
\label{fig:etabar_600}
\end{figure}

\section{Comparison with viscous hydrodynamics}
\label{sect:viscous}

Having obtained the exact solution of the Boltzmann equation one can compare the results with those obtained by first- and second-order viscous hydrodynamics approximations. 
It is well known that the first-order hydrodynamic theory suffers from conceptual difficulties. 
Nevertheless, we discuss this formulation below in order to analyze the system's behavior close to equilibrium and to empirically establish the correct relation between the shear viscosity and relaxation time.
In Sec.~\ref{sect:second} we introduce and discuss the equations of second-order viscous hydrodynamics. 

\subsection{First-order viscous hydrodynamics}
\label{sect:first}

At first-order the hydrodynamic equations for our simplified physical system reduce to the following two equations (see, for example, Eqs. (146) and (147) in \cite{Muronga:2003ta})
\begin{eqnarray}
\frac{d{\cal E}}{d\tau} = - \frac{{\cal E} + {\cal P}_{\rm eq}}{\tau} + \frac{4 \eta}{3\tau^2} 
\label{visc_fo1} 
\end{eqnarray}
and
\begin{eqnarray}
\frac{d{\cal S}_{\rm eq}}{d\tau} + \frac{{\cal S}_{\rm eq}}{\tau} = \frac{4 \eta}{3\tau^2 T},
\label{visc_fo2}
\end{eqnarray}
where $\eta$ is the shear viscosity. Since the value of the equilibrium energy density ${\cal E}_{\rm eq}$ is always equal to the non-equilibrium value ${\cal E}$ by construction, we have identified these two quantities. The equilibrium pressure and entropy density are defined through the thermodynamic relations
\begin{eqnarray}
{\cal P}_{\rm eq} = \frac{1}{3} {\cal E}_{\rm eq}, \quad {\cal E}_{\rm eq}+ {\cal P}_{\rm eq} = T {\cal S}_{\rm eq} \, ,
\label{visc_fo3}
\end{eqnarray}
where we have made use of the fact that the system obeys an ideal equation of state.
Combining Eqs.~(\ref{enmomcon2}) and (\ref{visc_fo1}) we conclude that
\begin{eqnarray}
{\cal P}_T = {\cal P}_{\rm eq} + \frac{\Pi}{2}, \quad
{\cal P}_L = {\cal P}_{\rm eq} - \Pi 
\label{eq:ptpl}
\end{eqnarray}
and
\begin{eqnarray}
\Pi = \frac{2}{3} ({\cal P}_T-{\cal P}_L) = \frac{4\eta}{3\tau}.
\label{Pi-fo}
\end{eqnarray}
The quantity $\Pi$ is the rapidity-rapidity component of the shear tensor $\pi^{\mu\nu}$ \cite{Muronga:2003ta} and its magnitude measures deviations of the energy-momentum tensor from the perfect-fluid form
\begin{eqnarray}
\pi^{\mu\nu} \pi_{\mu\nu} = \frac{3}{2} \Pi^2.
\label{pimunu2}
\end{eqnarray}

Using Eq.~(\ref{LM1}) one finds that both (\ref{visc_fo1}) and (\ref{visc_fo2}) lead to the same equation for the temperature, namely
\begin{eqnarray}
\frac{dT}{d\tau}+\frac{T}{3\tau}=\frac{4 {\bar \eta}_{\rm eff}}{9 \tau^2},
\label{eta-fo}
\end{eqnarray}
where ${\bar \eta}_{\rm eff}$ is the viscosity to entropy density ratio, ${\bar \eta}_{\rm eff} = \eta/{\cal S}_{\rm eq}$. 
Equation (\ref{eta-fo}) may be used to calculate ${\bar \eta}_{\rm eff}$ as a function of the proper time. 
This result may be compared with the actual value of ${\bar \eta}$ used to solve the kinetic equation (\ref{solT}). 
In this way one can check how much the first-order hydrodynamics is consistent with the results of the kinetic theory.  
One expects that it should only be reliable at large proper times and that in this limit ${\bar \eta}_{\rm eff}$ obtained from (\ref{eta-fo}) should converge to the true shear viscosity of the system. 

The results of this calculation are presented in Figs.~\ref{fig:etabar_300} and \ref{fig:etabar_600}.  
We compare the result with the two results from the literature (see, for example, Eqs. (8.78) and (8.89) from \cite{Cerc:2002} and use ${\cal P}_{\rm eq}/{\cal S}_{\rm eq} = T/4$)
\begin{eqnarray}
\eta = \frac{2}{3} {\cal P}_{\rm eq} \tau_{\rm eq},
\quad 
{\bar \eta} = \frac{T \tau_{\rm eq}}{6}, \label{dif-etabar6} \\
\eta = \frac{4}{5} {\cal P}_{\rm eq} \tau_{\rm eq},
\quad
{\bar \eta} = \frac{T \tau_{\rm eq}}{5}.
\label{dif-etabar5}
\end{eqnarray}
Clearly, our numerical study favors  Eq.~(\ref{dif-etabar5}). 
To the best of our knowledge,  Eq.~(\ref{dif-etabar5}) was first derived in \cite{Anderson1974466} and then reproduced in \cite{Czyz:1986mr}, where the complete set of the kinetic coefficients for the quark-antiquark plasma, including the color conductivity coefficient, has been derived.\footnote{Note that  \cite{Anderson1974466} and \cite{Czyz:1986mr} use the units where $h=1$ and the calculations are done for classical statistics. The result \mbox{$\eta = 4 T^4 \tau_{\rm eq}/(5 \pi^2)$} (for one internal degree of freedom) is obtained from Eq. (74) in \cite{Anderson1974466} by taking the ultrarelativistic limit and dividing by $8 \pi^3$. The same result is obtained for $\eta$ if Eq. (6.12) in \cite{Czyz:1986mr} is divided by $16 \pi^3$. The extra factor of 2 is needed, since both quarks and antiquarks are considered in \cite{Czyz:1986mr}.} Recently, the result (\ref{dif-etabar5}) has been also obtained, among many other results, in Ref. ~\cite{Romatschke:2011qp}.

The numerical results shown in Figs.~\ref{fig:etabar_300} and \ref{fig:etabar_600} empirically demonstrate that the correct relationship between the shear viscosity and the relaxation time is ${\bar \eta} = T \tau_{\rm eq}/5$.
We note that if one uses the Grad-Israel-Stewart approximation truncated at second order in moments one erroneously obtains ${\bar \eta} = T \tau_{\rm eq}/6$ \cite{Romatschke:PC}.  
If one instead uses the Chapman-Enskog method \cite{Anderson1974466,Romatschke:2011qp}, a complete second order Grad expansion \cite{Denicol:2012cn}, or asymptotic expansion without moment expansion, one obtains the correct value of ${\bar \eta} = T \tau_{\rm eq}/5$.
Whether one obtains ${\bar \eta} = T \tau_{\rm eq}/6$ or ${\bar \eta} = T \tau_{\rm eq}/5$ is not specific to second-order viscous hydrodynamics, but instead is a result of the approximations used when treating the collisional kernel itself.  
We return to this issue in Sec.~\ref{sect:latetime} where we employ a late time expansion of the kinetic solution, viscous hydrodynamics, and anisotropic hydrodynamics without moment expansion.  
In all cases studied one finds ${\bar \eta} = T \tau_{\rm eq}/5$.

\subsection{Second-order viscous hydrodynamics}
\label{sect:second}

In second-order viscous hydrodynamics the system's dynamics is described by the energy evolution equation supplemented by the shear viscous stress evolution equation (see, for example, Eqs. (175) and (178) in \cite{Muronga:2003ta})
\begin{eqnarray}
\partial_\tau {\cal E}&=&-\frac{{\cal E}+{\cal P}}{\tau}+\frac{\Pi}{\tau} \; ,
\nonumber \\
\partial_\tau\Pi&=&-\frac{\Pi}{\tau_\pi}+\frac{4}{3}\frac{\eta}{\tau_\pi\tau}-\beta\frac{\Pi}{\tau}\,,
\label{eq:vhydro}
\end{eqnarray}
where $\tau_\pi = 5 \bar\eta/T$ is the shear relaxation time. 
Viscous hydrodynamics practitioners most often use $\beta = 4/3$ which we will refer to as the Israel-Stewart (IS) prescription.  
We will also compare the exact solutions with the complete second-order treatment from Ref.~\cite{Denicol:2012cn} which, within RTA, gives $\beta = 38/21$.  
We will refer to the second choice as the DNMR prescription.\footnote{Reference~\cite{Jaiswal:2013npa} has also obtained $\lambda = 38/21$ with a different technique.}  In both cases one can compute the transverse and
longitudinal pressures using Eq.~(\ref{eq:ptpl}). 
To be consistent with the exact solution and the anisotropic hydrodynamics approximation we assume an ideal equation of state for the viscous hydrodynamical approximations. 
The results obtained from (\ref{eq:vhydro}) will be compared with the exact solutions together with anisotropic hydrodynamics results in the next Section.\footnote{We note that in the conformal limit, it is now standard to include an additional term proportional to $\Pi^2$ in the dynamical equation for $\Pi$ \cite{Romatschke:2009im}.  In App.~\ref{app:lambda1} we assess the affect of such a term.}

\section{Comparison with anisotropic hydrodynamics}
\label{sect:aniso}

\begin{figure*}[t]
\begin{center}
\includegraphics[angle=0,width=1.0\textwidth]{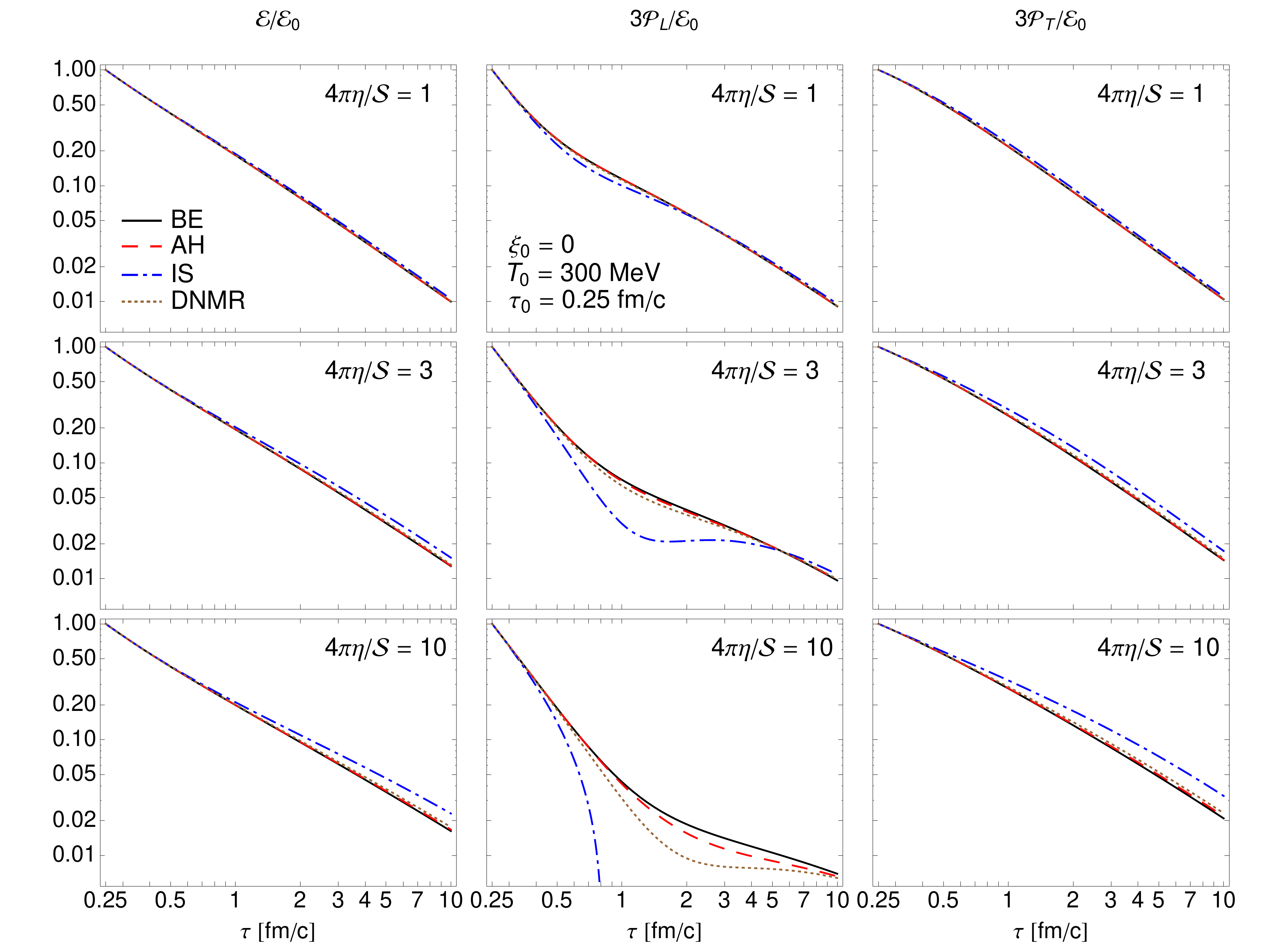}
\end{center}
\caption{(Color online) Time dependence of the energy density, the longitudinal pressure, and the transverse pressure (panels from left to right, respectively). The black solid, red dashed, blue dashed-dotted, and brown dotted lines describe the results obtained from the kinetic equation, anisotropic hydrodynamics, Israel-Stewart theory, and DNMR approach, respectively. The initial conditions in this figure are $T_0$ = 300 MeV and $\xi_0=0$.
}
\label{fig:PLPTE_300_0}
\end{figure*}

\begin{figure*}[t]
\begin{center}
\includegraphics[angle=0,width=1\textwidth]{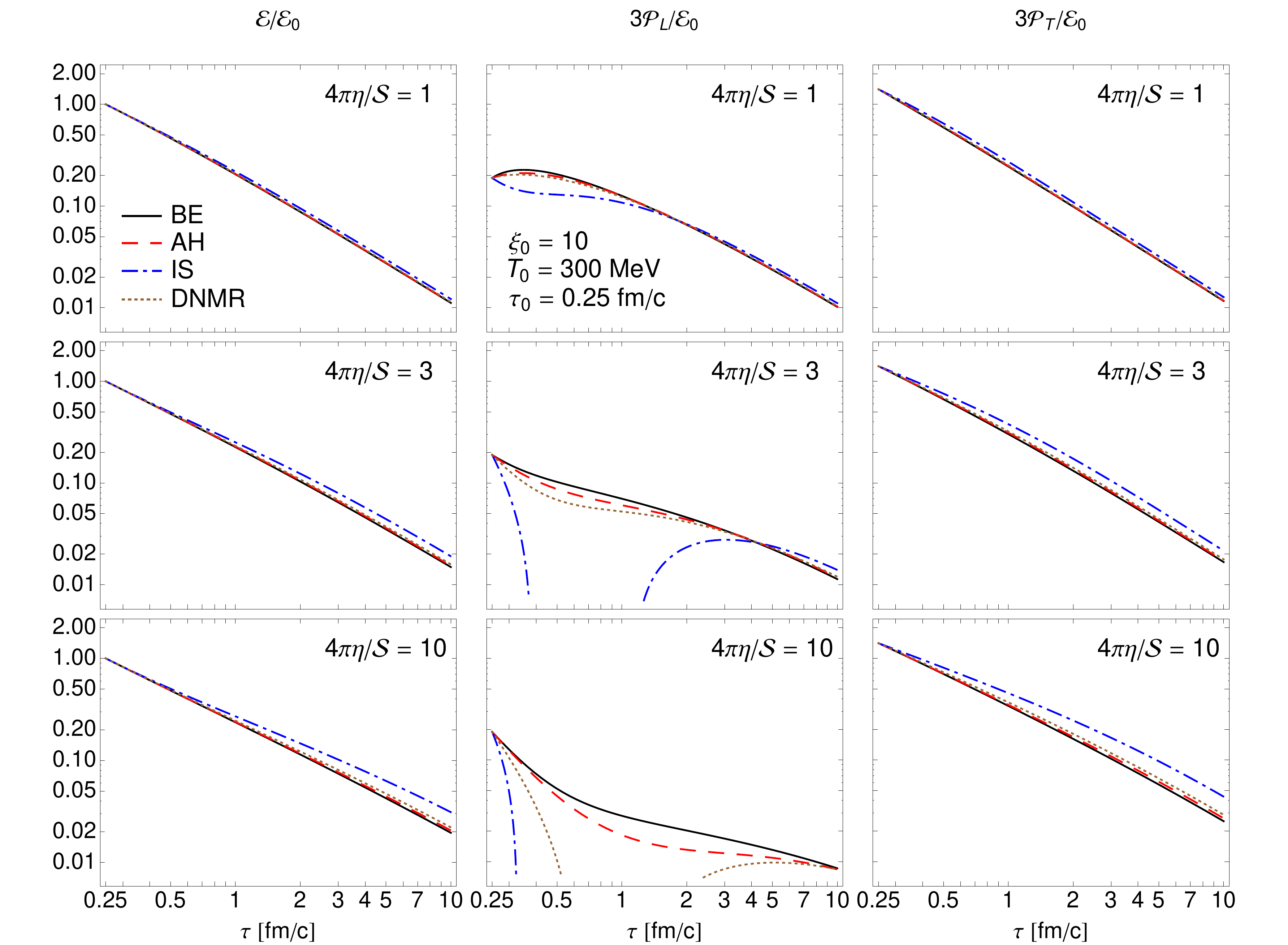}
\end{center}
\caption{(Color online) Same as Fig. \ref{fig:PLPTE_300_0} but for $T_0$ = 300 MeV and $\xi_0=10$.
}
\label{fig:PLPTE_300_10}
\end{figure*}

\begin{figure*}[t]
\begin{center}
\includegraphics[angle=0,width=1\textwidth]{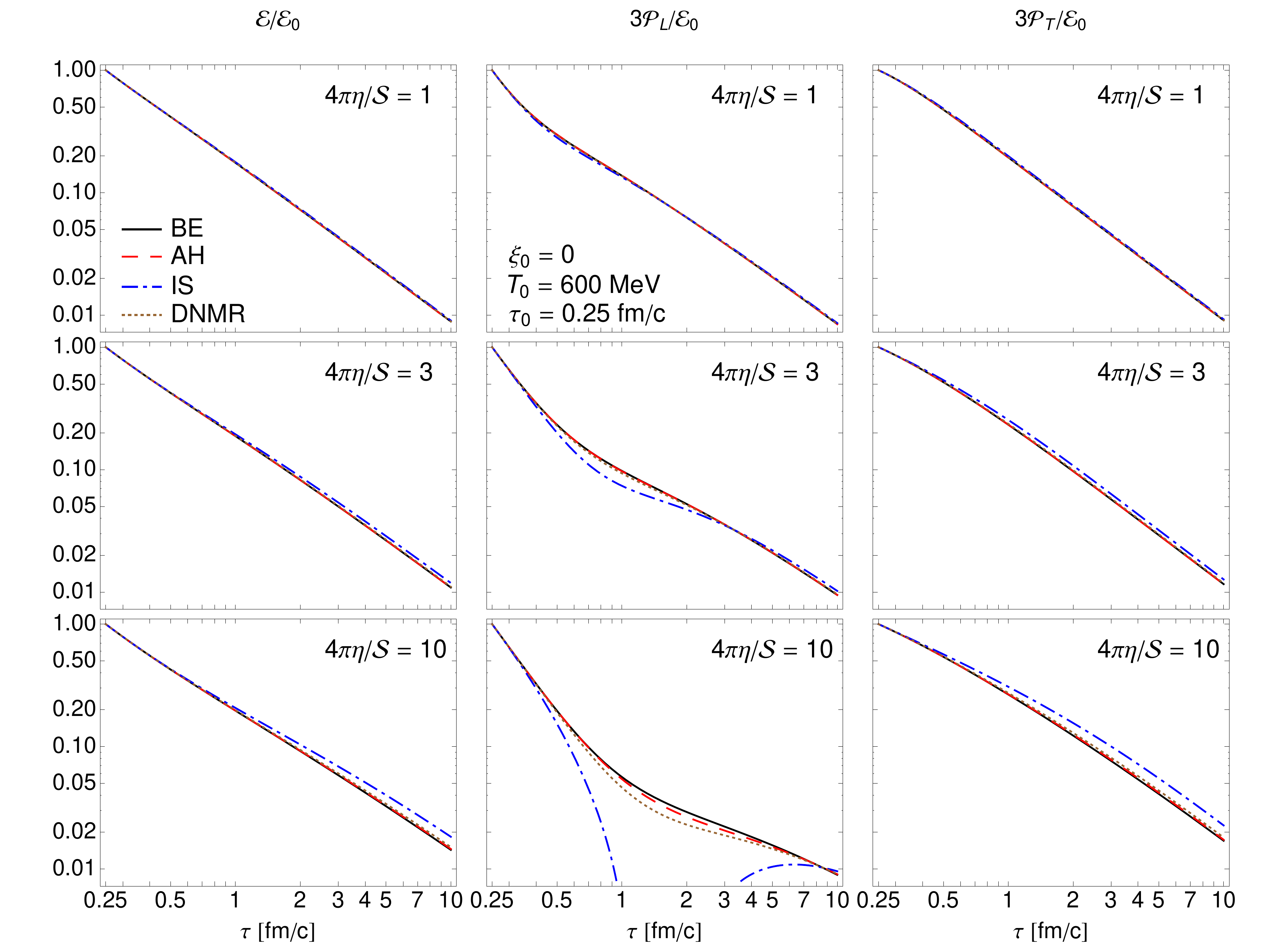}
\end{center}
\caption{(Color online) Same as Fig. \ref{fig:PLPTE_300_0} but for $T_0$ = 600 MeV and $\xi_0=0$. .
}
\label{fig:PLPTE_600_0}
\end{figure*}

\begin{figure*}[t]
\begin{center}
\includegraphics[angle=0,width=1\textwidth]{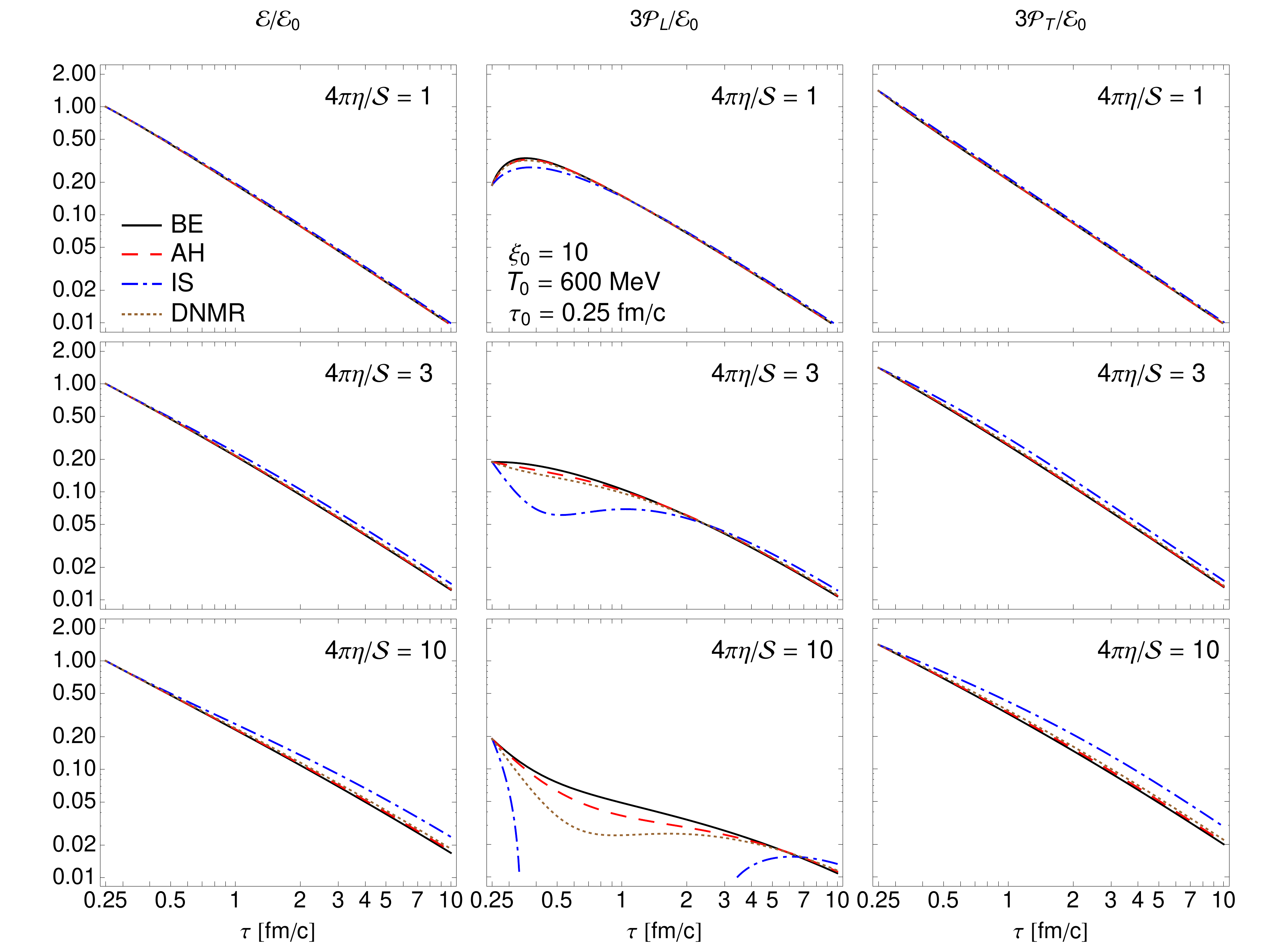}
\end{center}
\caption{(Color online) Same as Fig. \ref{fig:PLPTE_300_0} but for $T_0$ = 600 MeV and $\xi_0=10$. .
}
\label{fig:PLPTE_600_10}
\end{figure*}

We now turn to the comparison of our exact solutions of the kinetic equation (\ref{kineq}) with those obtained using the anisotropic hydrodynamic approximation \cite{Florkowski:2010cf,Martinez:2010sc,Ryblewski:2010bs,Martinez:2010sd,Ryblewski:2011aq,Martinez:2012tu,Ryblewski:2012rr,Ryblewski-0954-3899-40-9-093101,Florkowski:2012ax,Florkowski:2012as}. 
The anisotropic hydrodynamics framework is based on the analysis of the zeroth and first moments of the kinetic equation \cite{Martinez:2010sc,Martinez:2010sd,Martinez:2012tu}.  
In this approximation one assumes that to leading order the distribution function is given by a spheroidal Romatschke-Strickland form (RSF) \cite{Romatschke:2003ms} defined by the two time-dependent parameters: the transverse momentum scale $\Lambda(\tau)$ and the anisotropy parameter $\xi(\tau)$ \cite{Martinez:2010sc}.
All physical quantities may be expressed in terms of $\Lambda(\tau)$ and $\xi(\tau)$. 

For the case of an RSF obtained by the modification (stretching or squeezing) of an isotropic Boltzmann distribution, the energy density can be expressed as
\begin{eqnarray}
{\cal E} = \frac{6 g_0 \Lambda^4}{\pi^2} {\cal R}(\xi) \, ,
\label{epsahydro}
\end{eqnarray}
and the parton number density is
\begin{eqnarray}
n = \frac{2 g_0 \Lambda^3}{\pi^2 \sqrt{1+\xi}} \, .
\label{nahydro}
\end{eqnarray}
Similarly, the two pressures are obtained from the expressions
\begin{eqnarray}
{\cal P}_T = \frac{3g_0 \Lambda^4}{\pi^2} {\cal R}_T(\xi) \, , \quad 
{\cal P}_L = \frac{3g_0 \Lambda^4}{\pi^2} {\cal R}_L(\xi) \, ,
\label{PTPLinMS}
\end{eqnarray}
where the various ${\cal R}$ functions are defined in App.~\ref{app:H}.
The RSF entropy density in this case equals
\begin{eqnarray}
{\cal S} = 4 n = \frac{8 g_0 \Lambda^3}{\pi^2} \, .
\label{sigmainMS}
\end{eqnarray}
We recall that $g_0$ is the degeneracy factor accounting for all internal degrees of freedom except for spin. 

From the zeroth and first moment of the Boltzmann equation one obtains two dynamical equations \cite{Martinez:2010sc}
\begin{equation}
\frac{\partial_\tau\xi}{1+\xi} = \frac{2}{\tau} -  \, \frac{4 {\cal R}(\xi)}{\tau_{\rm eq}^{\rm AH}} \,  \, \frac{{\cal R}^{3/4}(\xi)\sqrt{1+\xi}-1}{2 {\cal R}(\xi) + 3 (1+\xi) {\cal R}'(\xi)} \, ,
\label{MS1}
\end{equation}
and
\begin{equation}
\frac{1}{1+\xi}\frac{\partial_\tau\Lambda}{\Lambda}  = \frac{{\cal R}'(\xi) }{\tau_{\rm eq}^{\rm AH}} \, \, \frac{{\cal R}^{3/4}(\xi)\sqrt{1+\xi}-1}{2 {\cal R}(\xi) + 3 (1+\xi) {\cal R}'(\xi)} \, ,
\label{MS2}
\end{equation}
where $\tau_{\rm eq}^{\rm AH}$ is the anisotropic hydrodynamics relaxation time. 
We solve Eqs.~(\ref{MS1}) and (\ref{MS2}) with initial conditions which are exactly the same as in the numerical calculations of the kinetic equation described in the previous Sections. 
This is possible since the initial conditions for the kinetic equation were chosen to have the same functional form. 

In Eqs.~(\ref{MS1}) and (\ref{MS2}) we allow the relaxation time $\tau_{\rm eq}^{\rm AH}$ to be different from the relaxation time $\tau_{\rm eq}$ used in the original kinetic equation (\ref{kineq}).  
In fact, as we will demonstrate in Sec.~\ref{sect:latetime}, by making asymptotic expansions of the anisotropic hydrodynamics equations (\ref{MS1}) and (\ref{MS2}) and the kinetic equation (\ref{solT}) one finds that 
\begin{eqnarray}
\tau_{\rm eq}^{\rm AH} 
= \frac{\tau_{\rm eq}}{2} \, , \quad |\xi| \ll 1 \, .
\label{teq2}
\end{eqnarray} 
A simple argument why (\ref{teq2}) should hold is the following: In Ref.~\cite{Martinez:2010sc} the matching between anisotropic hydrodynamics and the Israel-Stewart theory has been made in the case of small anisotropies. 
This matching leads to the formula
\begin{eqnarray}
\Gamma \equiv
 \frac{1}{\tau_{\rm eq}^{\rm AH} }
=\frac{2}{\tau_\pi} =
\frac{8 \,{\cal P}_{\rm eq}}{5 \eta}.
\label{matching}
\end{eqnarray}
In view of our results presented in Sec.~\ref{sect:first} we know that the correct value of the viscosity is $\eta = 4 {\cal P}_{\rm eq} \tau_{\rm eq}/5 $, see Eq.~(\ref{dif-etabar5}). 
Hence, Eq.~(\ref{matching}) leads to $\tau_{\rm eq}=\tau_\pi$ and (\ref{teq2}). 
If we kept the relaxation times $\tau_{\rm eq}^{\rm AH}$ and $\tau_{\rm eq}$ equal, the system described by anisotropic hydrodynamics would have a shear viscosity which is two times larger than the viscosity found in the exact solution. Hence one must adjust $\tau_{\rm eq}^{\rm AH}$ by a factor of two.
A formal proof of (\ref{teq2}) is given in the next Section. 

If the system is off equilibrium, the proper matching between $\tau_{\rm eq}^{\rm AH}$ and $\tau_{\rm eq}$ is more difficult to find. The numerical analyses of the solutions indicates that
\begin{eqnarray}
 \tau_{\rm eq}^{\rm AH} 
= \frac{T}{2 \Lambda} \, \tau_{\rm eq} \, ,
\label{teq2TL1}
\end{eqnarray} 
or, equivalently
\begin{eqnarray}
 \tau_{\rm eq}^{\rm AH} 
= \frac{5 {\bar \eta}}{2 \Lambda}.
\label{teq2TL2}
\end{eqnarray} 
In remainder of this Section we use Eqs. (\ref{dif-etabar5}) and (\ref{teq2TL2}) and present comparisons between the exact solutions of the kinetic equation, the results of the two second-order viscous hydrodynamics approximations, and the results of the anisotropic hydrodynamics approximation.
We note that the above prescription is different than the original Martinez-Strickland prescription \cite{Martinez:2010sc} for the relaxation time, which results in $\tau_{\rm eq}^{\rm MS} = \tau_{\rm eq}/2$.  We find in practice that the $\tau_{\rm eq}^{\rm AH}$ prescription given by Eq.~(\ref{teq2TL1}) (or equivalently Eq.~(\ref{teq2TL2})) results in much better agreement between anisotropic hydrodynamics and the exact kinetic solution.

In Figs.~\ref{fig:PLPTE_300_0}$\,-\,$\ref{fig:PLPTE_600_10} we show the time dependence of the energy density, the longitudinal pressure, and the transverse pressure (three columns of panels from left to right, respectively) obtained for three different values of the viscosity: 
$4\pi{\bar \eta} = 1, 3, 10$ (three rows of panels from top to bottom, respectively). 
The energy density is normalized to its initial value, while the longitudinal and transverse pressures are normalized to one third of the initial energy density. 
In this way, the late-time behavior of the displayed quantities becomes similar. 
Figures~\ref{fig:PLPTE_300_0}$\,-\,$\ref{fig:PLPTE_600_10} differ in the choice of initial conditions. 
We use $T_0$ = 300 MeV and $\xi_0=0$ in Fig.~\ref{fig:PLPTE_300_0}. 
The consecutive figures show the cases: $T_0$ = 300 MeV and $\xi_0=10$, $T_0$ = 600 MeV and $\xi_0=0$, and $T_0$ = 600 MeV and $\xi_0=10$. 
The black solid, red dashed, blue dashed-dotted, and brown dotted lines in Figs.~\ref{fig:PLPTE_300_0}$\,-\,$\ref{fig:PLPTE_600_10} are the results obtained from the kinetic equation, anisotropic hydrodynamics, Israel-Stewart theory, and the DNMR approach, respectively.

In all cases considered one observes noticeable differences between the exact results and the standard Israel-Stewart approximation. 
If the shear viscosity becomes large, the Israel-Stewart theory results in negative longitudinal pressure.\footnote{In Figs.~\ref{fig:PLPTE_300_0}$\,-\,$\ref{fig:PLPTE_600_10}, in order to more precisely compare the various approximations, we have used a logarithmic scale for the horizontal and vertical axes.  With this scaling for the vertical axis, negative values of the longitudinal pressure lead to regions where the logarithm is undefined.}
In all cases studied, the agreement between viscous hydrodynamics and kinetic theory is dramatically improved if one uses the DNMR approximation and is further improved if one uses the anisotropic hydrodynamics approximation.  We note, however, that the problem of negative pressure, although lessened somewhat, still exists in the DNMR approach, as can be seen from the bottom middle panel of Fig.~\ref{fig:PLPTE_300_10}.

\begin{figure}[t]
\begin{center}
\includegraphics[angle=0,width=0.95\textwidth]{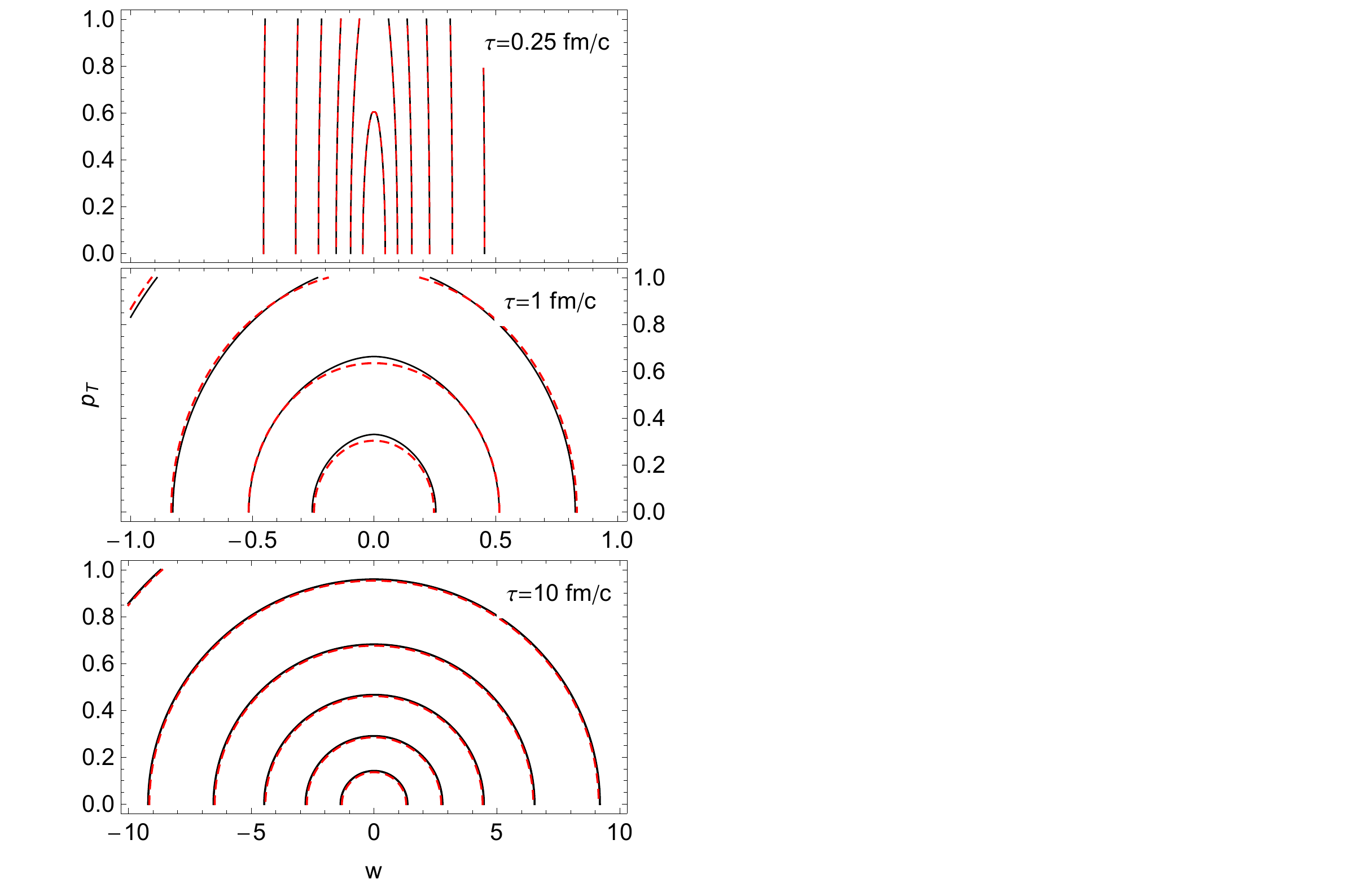}
\end{center}
\caption{(Color online) Distribution function contours obtained from the exact kinetic solution (black lines) and the anisotropic hydrodynamics approximation (red dashed lines) for $T_0$ = 600 MeV, $\xi_0=10$, and $4\pi {\bar \eta} = 1$.}
\label{fig:G_600_10_1}
\end{figure}

\begin{figure}[t]
\begin{center}
\includegraphics[angle=0,width=0.95\textwidth]{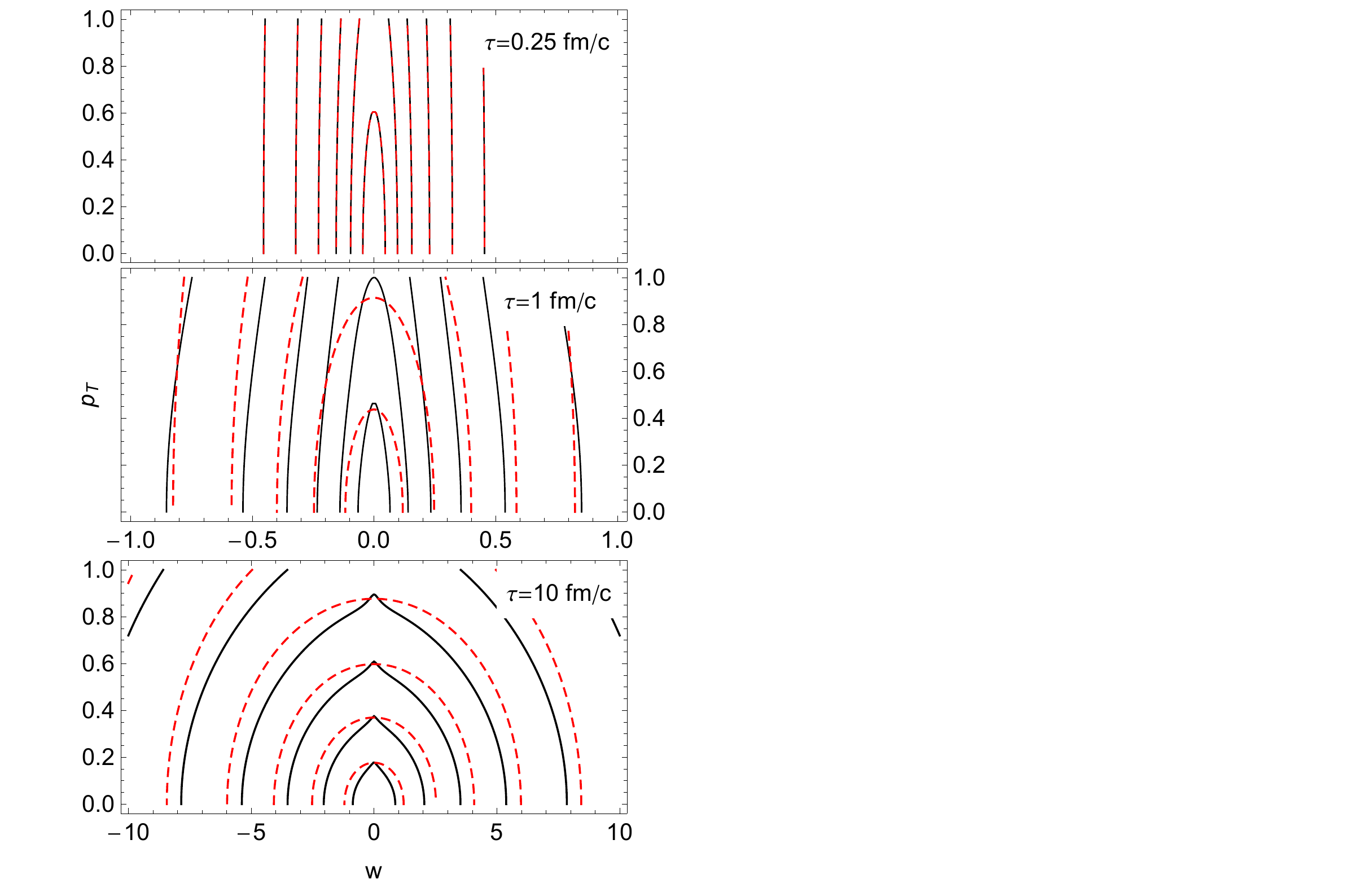}
\end{center}
\caption{(Color online) Distribution function contours obtained from the exact kinetic solution (black lines) and the anisotropic hydrodynamics approximation (red dashed lines) for $T_0$ = 600 MeV, $\xi_0=10$, and $4\pi {\bar \eta} = 10$.}
\label{fig:G_600_10_10}
\end{figure}

Once the evolution of the effective temperature is known via Eq.~(\ref{solT}) this can be used in Eq.~(\ref{solG}) to determine the exact evolution of the distribution function in momentum space.  
This allows one to extract more detailed information than allowed by the moments of the distribution alone.
In Figs.~\ref{fig:G_600_10_1}$\,-\,$\ref{fig:G_600_10_10} we compare contour plots of the distribution function obtained with the exact kinetic solution (black lines) and the anisotropic hydrodynamics approximation (red dashed lines). 
We see from these figures that at early times corresponding to $\tau \sim$ 1 fm/c there is disagreement between anisotropic hydrodynamics and the exact solution for the distribution function.
One can see, particularly in Fig.~\ref{fig:G_600_10_10} which presents the case $4 \pi \bar\eta = 10$, that the distribution function is not spheroidal.  
In fact, one can see what appears to be a superposition of two spheroids, one which is governed by pure free-streaming evolution coming from the first term in Eq.~(\ref{solG}) and the second coming from an equilibrating component coming from the second term in Eq.~(\ref{solG}).  
This suggests that it may be more accurate to use a form which is linear superposition of two spheroids.  We leave this possibility for future work.

\begin{figure}[t]
\begin{center}
\includegraphics[angle=0,width=0.95\textwidth]{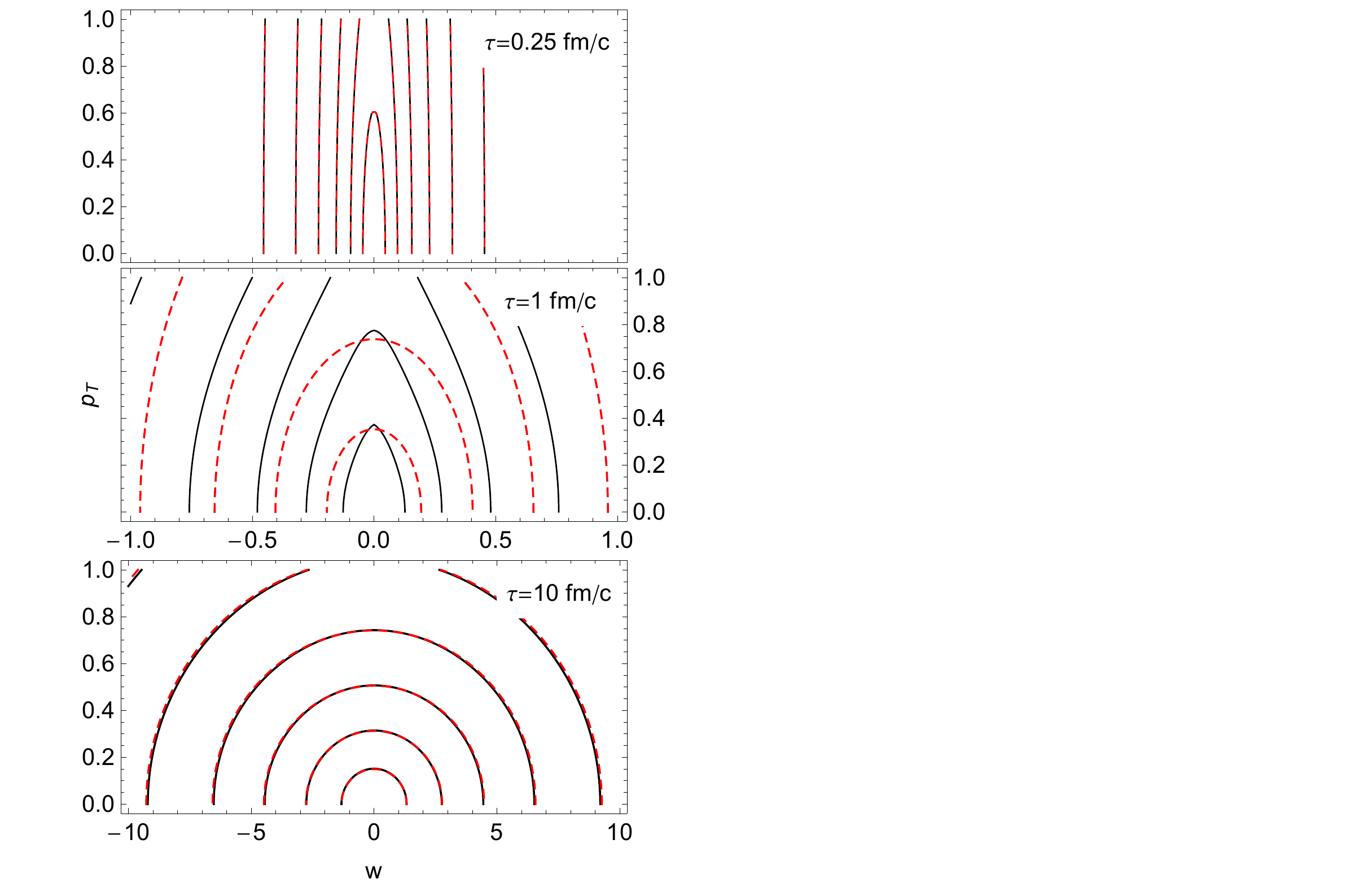}
\end{center}
\caption{(Color online) Distribution function contours obtained from the exact kinetic solution (black lines) and the anisotropic hydrodynamics approximation (red dashed lines) for $T_0$ = 600 MeV, $\xi_0=10$, and $4\pi {\bar \eta} = 3$.}
\label{fig:G_600_10_3}
\end{figure}

\section{Late-time behavior}
\label{sect:latetime}

In this Section we analyze the late-time behavior of the system described by anisotropic hydrodynamics, according to Eqs.~(\ref{MS1}) and (\ref{MS2}), by the kinetic equation (\ref{kineq}), and by the second-order viscous hydrodynamics equations (\ref{eq:vhydro}). We prove that the relation~(\ref{teq2}) should indeed be satisfied in order to achieve the agreement between anisotropic hydrodynamics and the kinetic theory and
that ${\bar \eta} = T \tau_{\rm eq}/5$ is the correct relationship between the shear viscosity and the relaxation time in the near-equilibrium limit.

\subsection{Asymptotic expansion of anisotropic hydrodynamics}

We start with Eqs.~(\ref{MS1}) and (\ref{MS2}). Since at late times $\xi \rightarrow 0$, we can linearize these two equations in $\xi$. Treating $\xi$ and $\partial_\tau \xi$ as order $\epsilon$ we expand the first equation to order $\epsilon^2$ to obtain
\begin{equation}
\partial_\tau\xi = \frac{2}{\tau} + \left( \frac{2}{\tau} - \frac{\Gamma}{2} \right) \xi
- \frac{17}{63} \Gamma \xi^2 
+ {\cal O}(\xi^3) \, .
\label{eq:xiexp}
\end{equation}
Here $\Gamma$ is defined by (\ref{matching}). Similarly, we can expand the second equation to obtain
\begin{equation}
\partial_\tau\Lambda = -\frac{1}{12} \Gamma \Lambda \xi  + \frac{187}{3780} \Gamma \Lambda \xi^2 + {\cal O}(\xi^3) \, .
\label{eq:lambdaexp}
\end{equation}
In the next step, we find the solution for $\xi$.  One finds empirically that $\xi$ decays like
\begin{equation}
\lim_{\tau \rightarrow \infty} \xi = \frac{A}{\tau} + \frac{B}{\tau^2} + {\cal O}\left(\frac{1}{\tau^3}\right) .
\end{equation}
Plugging this form into (\ref{eq:xiexp}) and matching terms of order $\tau^{-1}$ and $\tau^{-2}$ on the left and right hand sides one finds
\begin{equation}
\lim_{\tau \rightarrow \infty} \xi(\tau) = \frac{4}{\Gamma\tau} + \frac{968}{63 (\Gamma \tau)^2} + {\cal O}\left(\frac{1}{\tau^3}\right) .
\end{equation}
Inserting this solution on the right hand side of (\ref{eq:lambdaexp}) and expanding through ${\cal O}(\tau^{-2})$ one obtains
\begin{equation}
\lim_{\tau \rightarrow \infty} \frac{1}{\Lambda} \partial_\tau\Lambda = -\frac{1}{3} - \frac{22}{45} \frac{1}{\Gamma \tau^2} 
+ {\cal O}(\tau^{-3}) \, .
\label{eq:lambdaexp2}
\end{equation}
Solving this differential equation and taking the limit $\Gamma \tau \gg 1$ we obtain
\begin{equation}
\lim_{\tau \rightarrow \infty} \Lambda(\tau) = \frac{C}{\tau^{1/3}} \left( 1 + \frac{22}{45} \frac{1}{\Gamma \tau} + {\cal O}\left(\tau^{-2}\right) \right) , 
\end{equation}
where $C$ is an undetermined constant.  Having determined the asymptotic behavior of $\xi$ and $\Lambda$ we can now determine the asymptotic expansion of the energy density ${\cal E} = {\cal R}(\xi) {\cal E}_{\rm eq}(\Lambda)$ 
\begin{equation}
\lim_{\tau \rightarrow \infty} {\cal E}(\tau) = \frac{D}{\tau^{4/3}} \left( 1 - \frac{32}{45} \frac{1}{\Gamma \tau} + {\cal O}\left(\tau^{-2}\right) \right) .
\label{epsAH}
\end{equation}

\subsection{Asymptotic expansion of the relaxation-time-approximation integral equation}

The integral equation for the energy density is obtained from (\ref{solT})
\begin{eqnarray}
{\cal E}(\tau) &=& D(\tau,\tau_0) {\cal E}_0
\frac{{\cal H}\left(\frac{\tau_0}{\tau} x_o^{-1/2} \right)}{{\cal H}\left(x_o^{-1/2} \right)} \nonumber \\
&& + \int_{\tau_0}^{\tau} \frac{d\tau^\prime}{2\tau_{\rm eq}(\tau^\prime)} D(\tau,\tau^\prime) 
{\cal E}(\tau^\prime) {\cal H}\left(\frac{\tau^\prime}{\tau}\right) .
\label{eq:inteq}
\end{eqnarray}
We seek the large-$\tau$ asymptotic solution of this equation and once again search for
a solution of the form 
\begin{equation}
\lim_{\tau \rightarrow \infty} {\cal E}(\tau) = A\left(\frac{\tau_{\rm eq}}{\tau}\right)^{4/3} \left( 1 + B \frac{\tau_{\rm eq}}{\tau} + {\cal O}\left(\tau^{-2}\right) \right) .
\end{equation}
As $\tau \rightarrow \infty$ the first term in (\ref{eq:inteq}) goes to zero exponentially fast, so we can ignore it.  In order to evaluate the integral we recognize that the integral is dominated by the end of the integration region where $\tau^\prime \sim \tau$ due to the damping function $D$. As a result, we can proceed by expanding the ${\cal H}$ function in a power series around $\tau^\prime = \tau$ from below.  In order to extract the asymptotic coefficients necessary, it suffices to expand ${\cal H}$ to second order, see Appendix \ref{app:H},
\begin{equation}
\lim_{\tau^\prime \rightarrow \tau} 
{\cal H}\left( \frac{\tau^\prime}{\tau} \right) = 2 + \frac{8(\tau^\prime - \tau)}{3\tau}
+ \frac{4(\tau^\prime - \tau)^2}{5\tau^2} + {\cal O}\left((\tau^\prime - \tau)^3\right) .
\end{equation}
Inserting the asymptotic expansion above on the left and right hand sides of Eq.~(\ref{eq:inteq}), performing the integral on the right hand side, and discarding terms which go to zero exponentially in $\tau - \tau_0$ one obtains
\begin{eqnarray}
&& A\left(\frac{\tau_{\rm eq}}{\tau}\right)^{4/3} \left( 1 + B \frac{\tau_{\rm eq}}{\tau} \right) 
= A\left(\frac{\tau_{\rm eq}}{\tau}\right)^{4/3} \left( 1 + B \frac{\tau_{\rm eq}}{\tau} \right)
\nonumber \\
&& - \frac{A}{45} \left(\frac{\tau_{\rm eq}}{\tau}\right)^{10/3} (16 + 45 B) + {\cal O}\left(\tau^{-13/3}\right) .
\end{eqnarray}
Requiring equivalence between the left and right we obtain $B = -16/45$ giving
\begin{equation}
\lim_{\tau \rightarrow \infty} {\cal E}(\tau) = A\left(\frac{\tau_{\rm eq}}{\tau}\right)^{4/3} \left( 1 - \frac{16}{45} \frac{\tau_{\rm eq}}{\tau} + {\cal O}\left(\tau^{-2}\right) \right).
\label{epsKIN}
\end{equation}
Comparing (\ref{epsKIN}) with (\ref{epsAH}) one obtains (\ref{teq2}).  We note that one can find this result derived in a different manner in Ref.~\cite{Baym:1984np}.

\subsection{Asymptotic expansion of the second-order viscous hydrodynamics equations}
\label{subsec:vhydroasymp}

We start with the viscous hydrodynamical equations (\ref{eq:vhydro}).  To proceed, we assume that $\tau_{\rm eq}=\tau_\pi$ is held constant.  The shear viscosity and the relaxation time are related via
\begin{equation}
\eta = \alpha \tau_{\rm eq} T {\cal S}  = \frac{4}{3} \alpha \tau_{\rm eq} {\cal E} \, ,
\end{equation}
where $\alpha$ will be determined via asymptotic expansion and matching.  Note that we have assumed an ideal equation of state in the last equality.
With these assumptions one finds that the energy density and shear $\Pi$ have the following asymptotic expansions
\begin{eqnarray}
\lim_{\tau \rightarrow \infty} {\cal E}(\tau) &=& A \left(\frac{\tau_{\rm eq}}{\tau}\right)^{4/3} + B \left(\frac{\tau_{\rm eq}}{\tau}\right)^{7/3} + {\cal O}(\tau^{-10/3}) \, ,
\nonumber \\
\lim_{\tau \rightarrow \infty} \Pi(\tau) &=& C \tau^{-7/3} + {\cal O}(\tau^{-10/3}) \, .
\end{eqnarray}
Inserting these expansions and requiring that in the limit $\tau \rightarrow \infty$ the coefficient of the leading ${\cal O}(\tau^{-10/3})$ term in the first equation vanishes gives $B = -C$.  Requiring that the coefficient of the leading ${\cal O}(\tau^{-7/3})$ term in the second equation vanishes gives $C = 16 A \alpha/9$.  Putting these results together one obtains
\begin{equation}
\lim_{\tau \rightarrow \infty} {\cal E}(\tau) = A\left(\frac{\tau_{\rm eq}}{\tau}\right)^{4/3} \left( 1 - \frac{16\alpha}{9} \frac{\tau_{\rm eq}}{\tau} + {\cal O}\left(\tau^{-2}\right) \right) .
\label{eq:vhae}
\end{equation}
Matching Eqs.~(\ref{epsKIN}) and (\ref{eq:vhae}) one obtains $\alpha = 1/5$ independently of the coefficient $\beta$ which appears in the second-order equations.  This gives the desired relation $\bar\eta =  T\tau_{\rm eq}/5$.

\section{Conclusions}
\label{sect:conclusions}

In this paper we presented an exact solution to the 0+1d Boltzmann equation in the relaxation time approximation.  
Our solution is appropriate for systems with time-independent or time-dependent relaxation times.
From this solution we were able to obtain to arbitrary numerical accuracy the proper-time evolution of all relevant bulk properties of the system: the energy density, transverse and longitudinal pressures, number density, and entropy density.
We then compared the exact kinetic theory solution to the standard Israel-Stewart second-order viscous hydrodynamics approximation (IS), a complete second-order viscous hydrodynamics approximations (DNMR), and the anisotropic hydrodynamics approximation.
We performed comparisons of the energy density and pressures for two different initial temperatures, two different initial anisotropies, and three different values for the shear viscosity to entropy density ratio.  

Our results show that, among the different approximations considered, the standard IS approximation was the poorest approximation to the exact RTA solution.
Comparatively, the DNMR second-order viscous hydrodynamics approximation represented a significant improvement over the IS approximation; however, like the standard IS approximation the DNMR approximation can result in predictions of negative longitudinal pressure.  Finally, in all cases tested the anisotropic hydrodynamics approximation most accurately reproduced the exact RTA solution.  The relative success of anisotropic hydrodynamics in reproducing the exact results is somewhat surprising since the equations used were derived at LO in the anisotropic expansion, only taking into account a spheroidal functional form for the one-particle distribution function.

In the process we were able to establish that there exists a factor of two difference between the relaxation time in the anisotropic hydrodynamics approximation and the exact relaxation time.  Additionally, we determined that for best agreement with the exact kinetic solution, the scale in the far-from-equilibrium anisotropic dynamics relaxation time should be set by the transverse temperature $\Lambda$.  In the context of second-order viscous hydrodynamics we determined empirically and analytically that the correct relationship between the shear viscosity and the relaxation time is $\bar\eta = T \tau_{\rm eq}/5$.   

Of course, our exact solution is restricted to the 0+1d Boltzmann equation in RTA.  As a consequence, the conclusions stated above are strictly applicable only in this context.  It is not currently possible to make a general statement about the ranking of the relative errors of the various approximations.  That being said it is certainly nice to have one exactly solvable case that can be used to assess different approximation schemes.  Since the exact solution obtained is applicable for arbitrary shear viscosity to entropy ratio it can be used to assess the efficacy of different far-from-equilibrium approaches.  Looking forward, knowledge of the exact solution in this simple situation could prove useful in the development of more comprehensive far-from-equilibrium approximation schemes.

\begin{acknowledgments}
We thank G.~Denicol and P. Romatschke for discussions.  
This work was supported in part by the Polish National Science Center with decision No.~DEC-2012/06/A/ST2/00390.
R.R. was supported by a Polish National Science Center grant with decision No.~DEC-2012/07/D/ST2/02125 and the Foundation for Polish Science.
\end{acknowledgments}

\appendix

\section{The ${\cal H}$ and ${\cal R}$ functions}
\label{app:H}

The functions ${\cal H}$, ${\cal H}_L$, and ${\cal H}_T$ are defined by the integrals
\begin{eqnarray}
{\cal H}(y) &=& y \int\limits_0^\pi d\phi \,
\sin\phi \, \sqrt{y^2 \cos^2\phi+\sin^2\phi} \, ,
\nonumber \\
{\cal H}_L(y) &=& y^3 \int\limits_0^\pi d\phi \, \frac{\sin\phi \cos^2\phi}{
\sqrt{y^2 \cos^2\phi+\sin^2\phi}} \, ,
\nonumber \\
{\cal H}_T(y) &=& y \int\limits_0^\pi d\phi \, \frac{\sin^3\phi }{
\sqrt{y^2 \cos^2\phi+\sin^2\phi}} \, .
\label{Hs}
\end{eqnarray}
There are simple relations connecting ${\cal H}$, ${\cal H}_L$, and ${\cal H}_T$ with the functions ${\cal R}$, ${\cal R}_L$, and ${\cal R}_T$ defined in Ref.~\cite{Martinez:2010sc}, namely
\begin{eqnarray}
{\cal H}\left(\frac{1}{\sqrt{1+\xi}}\right) &=& 2 \, \cal{R}(\xi) \, ,
\nonumber \\
{\cal H}_L\left(\frac{1}{\sqrt{1+\xi}}\right) &=& \frac{2}{3} {\cal R}_L(\xi) \, ,
\nonumber \\
{\cal H}_T\left(\frac{1}{\sqrt{1+\xi}}\right) &=& \frac{4}{3} {\cal R}_T(\xi) \, .
\end{eqnarray}
In the region $0.5 \leq y \leq 1$ the functions ${\cal H}(y)$ are very well approximated by the expressions
\begin{eqnarray}
{\cal H}\left(y\right) &\simeq& 2 + \frac{8}{3} (y - 1) 
+ \frac{4}{5} (y - 1)^2 + {\cal O}\left((y - 1)^3\right) ,
\nonumber \\
{\cal H}_L\left(y\right) &\simeq& \frac{2}{3} + \frac{8}{5} (y - 1) + \frac{36}{35} (y - 1)^2 
\nonumber \\ && \hspace{5mm} + \frac{8}{315} (y - 1)^3 + {\cal O}\left((y - 1)^4\right) ,
\nonumber \\
{\cal H}_T\left(y\right) &\simeq& \frac{4}{3} + \frac{16}{15} (y - 1)) - \frac{8}{35} (y - 1)^2 
\nonumber \\ && \hspace{5mm} + \frac{16}{315} (y - 1)^3 + {\cal O}\left((y - 1)^4\right) .  
\end{eqnarray}

\section{Including the $\lambda_1$ conformal term}
\label{app:lambda1}

\begin{figure}[t]
\begin{center}
\includegraphics[angle=0,width=1.0\textwidth]{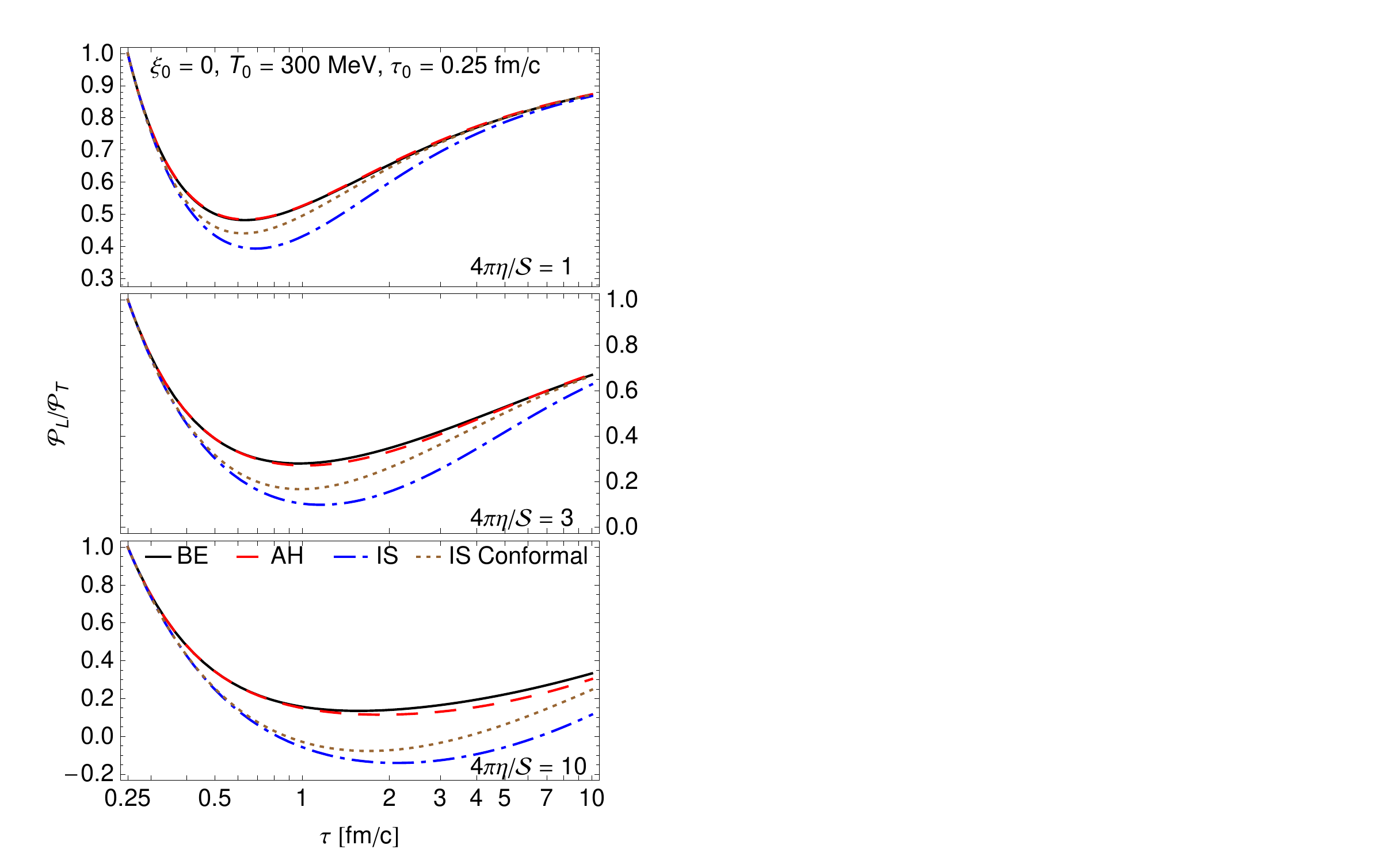}\end{center}
\caption{(Color online) Time dependence of the ratio of the longitudinal and transverse pressures for
$T_0$ = 300 MeV and $\xi_0=0$ at $\tau_0 =$ 0.25 fm/c.  
Shown in the plot are the exact kinetic
theory solution (black), the anisotropic hydrodynamics approximation (red dashed), the Israel-Stewart
equation without the conformal $\Pi^2$ term (blue dot-dashed), and the Israel-Stewart
equation with the conformal $\Pi^2$ term (brown dotted).  
Panels (top to bottom) show the
cases $4 \pi \bar\eta =$ 1, 3, and 10, respectively. 
In all cases we used ${\bar \eta} = T \tau_{\rm eq}/5$.  
}
\label{fig:IScomp}
\end{figure}

In this appendix we discuss the impact of including the full set of ``conformal'' second-order terms in the viscous hydrodynamical evolution.  
In the conformal limit one finds that an additional term is required in the second-order viscous hydrodynamical equations which is proportional to $\Pi^2$ \cite{Romatschke:2009im}
\begin{eqnarray}
\partial_\tau {\cal E}&=&-\frac{{\cal E}+{\cal P}}{\tau}+\frac{\Pi}{\tau} \; ,
\nonumber \\
\partial_\tau\Pi&=&-\frac{\Pi}{\tau_\pi}+\frac{4}{3}\frac{\eta}{\tau_\pi\tau}- \frac{4\Pi}{3\tau} 
-\frac{\lambda_1}{2\tau_\pi \eta^2} \Pi^2
\,,
\label{eq:vhydroConf}
\end{eqnarray}
where, in RTA, the coefficient $\lambda_1 = 5 \eta \tau_\pi/7$ \cite{Romatschke:2011qp,Ling:PC}.

In Fig.~\ref{fig:IScomp} we compare the solution of (\ref{eq:vhydroConf}) with and without the term proportional to $\Pi^2$ to the exact kinetic solution obtained via (\ref{solT}) and the anisotropic hydrodynamics approximation obtained via Eqs.~(\ref{MS1}) and (\ref{MS2}).  
As one can see from this figure, while the inclusion of the $\Pi^2$ does somewhat improve the agreement of the Israel-Stewart approximation with the exact solution, it still has a larger error associated with it than the anisotropic hydrodynamics approximation.  
In addition, we see in the bottom panel of Fig.~\ref{fig:IScomp} that the longitudinal pressure can become negative even when including the $\Pi^2$ term.

\vspace{6cm}

\bibliography{rtal}

\end{document}